\documentclass[aps,pra,twocolumn,showpacs,superscriptaddress]{revtex4-1}

\usepackage{amsfonts,mathrsfs,amsmath,amsthm,amssymb,bbold}
\usepackage[english]{babel}
\usepackage{float}
\usepackage{graphicx}
\usepackage{bm}
\usepackage{braket}
\usepackage{xcolor}
\usepackage{amsfonts}
\usepackage{comment}
\usepackage{dirtytalk}
\usepackage{cleveref}
\usepackage{url}
\usepackage{bbold}
\usepackage{realboxes}

\usepackage{enumitem}
\usepackage{natbib}
\usepackage{multirow}
\usepackage{tabularx}
\usepackage{textcomp}

\usepackage{float}

\usepackage{color, colortbl}
\definecolor{LightCyan}{rgb}{0.88,1,1}
\definecolor{Gray}{gray}{0.95}
\usepackage{array}
\usepackage{tabulary}
\newcolumntype{K}[1]{>{\centering\arraybackslash}m{#1}}

\usepackage{courier}

\newcommand{\tr}{\text{tr}}
\newcommand{\squote}{\textquotesingle}

\usepackage{listings}
\usepackage{xcolor}

\definecolor{bkgd}{RGB}{240,242,246}
\definecolor{ceruleanblue}{rgb}{0.16, 0.32, 0.75}
\definecolor{orange-red}{rgb}{1.0, 0.27, 0.0}
\definecolor{anotherblue}{RGB}{37,92,243}
\definecolor{blackblue}{RGB}{46,60,85}
\definecolor{goldyellow}{RGB}{199,146,12}
\lstdefinestyle{altstyle2}{
    backgroundcolor=\color{bkgd},
    basicstyle=\ttfamily\small\color{blackblue},
    breakatwhitespace=false,
    breaklines=true,
    captionpos=b,
    commentstyle=\color{goldyellow},
    keepspaces=true,
    keywordstyle=\color{orange-red},
    language=Python,
    numbersep=5pt,
    numberstyle=\tiny\color{ceruleanblue},
    showspaces=false,
    showstringspaces=false,
    showtabs=false,
    stringstyle=\color{anotherblue},
    tabsize=2
}
\newcounter{bla}

\begin{document}  
\title {$\mathtt{tqix.pis}$: A toolbox for quantum dynamics simulation of spin ensembles in Dicke basis
} 

\author{Nguyen Tan Viet}
\affiliation{FPT University, Hanoi, Vietnam}
\thanks{Electronic address: vietnthe153763@fpt.edu.vn}

\author{Nguyen Thi Chuong}
\affiliation{ILotusLand VietNam, Hochiminh City, Vietnam}
\author{Vu Thi Ngoc Huyen}
\affiliation{Center for Computational Materials Science, 
Institute for Materials Research, 
Tohoku University, Sendai, Miyagi 980-8577, Japan}
\author{Le Bin Ho}
\affiliation{Frontier Research Institute for Interdisciplinary Sciences, 
Tohoku University, Sendai 980-8578, Japan}
\affiliation{Department of Applied Physics, 
Graduate School of Engineering, 
Tohoku University, Sendai 980-8579, Japan}
\thanks{Electronic address: binho@fris.tohoku.ac.jp}


%

\date{\today}

\begin{abstract}
We introduce \texttt{tqix.pis}, 
a library of \texttt{tqix}, for
quantum dynamics simulation of
spin ensembles. 
The library emulates a dynamic process 
by a quantum circuit,
including initializing a quantum state, 
executing quantum operators, 
and measuring the final state. 
It utilizes collective processes 
in spin ensembles
to reduce the dimension from 
exponentially to quadratically
with the number of particles, 
i.e., the quantum state spans in Dicke basis.
It also facilitates the simulation time 
with multi-core processors 
and Graphics Processing Units.
The library is thus applicable 
for the simulation of ensembles of 
large number of particles
that have collective properties.
Various phenomena, 
such as spin squeezing, 
variational quantum squeezing, 
quantum phase transition,
and many-body quantum dynamics,
can be simulated using the library. \\

\noindent Note for installation:
Download the source code and run:
\texttt{pip3 install .}\\
Or install from PyPI, run: 
\texttt{pip3 install tqix}\\
%
Official website: https://vqisinfo.wixsite.com/tqix\\
https://tqix-developers.readthedocs.io/en/latest/index.html\\
https://pypi.org/project/tqix/\\

\end{abstract}
%
%

\maketitle

\section{Introduction}
\label{sec1}
The toolbox 
\texttt{tqix} is 
a Python-based simulation package for quantum measurement and
applications in quantum metrology and quantum tomography \cite{HO2021107902}. The package allows for designing quantum systems (state and operators) and performing quantum measurements. It embeds various conventional measurement operators, including Pauli, Stoke, MUB-POVM, and SIC-POVM,
with two backends to simulate the measurement results.
The applications in quantum metrology and quantum tomography were reported so far.

In this version, we introduce a new library 
called \texttt{tqix.pis} 
for programmable quantum dynamics
of spin ensembles with collective processes. 
In the collective processes, 
an $N$-particle state spans
in Dicke basis with $O(N^2)$-dimension,
smaller than the product $2^N$-dimension
\cite{PhysRev.93.99}. 
This is the backbone for the simulation 
of a large number of particles
with low computation costs.
Thus, the library is restricted 
to the simulation of collective 
phenomena in ensemble 
systems with collective processes
\cite{PhysRevA.78.052101},
which is different from a general simulation method,
where every particle (qubit) is accessible. 

The library executes a dynamic simulation
by a quantum circuit with 
a sequence of initialization collective state,
the action of collective operators,
and measurements in the Dicke basis. 
It supports 
parallelizing multi-core processors 
and Graphics Processing Units (GPUs)
to facilitate the running time.
In terms of application, 
it is suitable for studying spin squeezing \cite{MA201189},
variational quantum squeezing
\cite{PhysRevLett.123.260505,PhysRevX.11.041045},
quantum phase transition
\cite{sachdev_2011}, 
many-body quantum dynamics
\cite{Smith2019},
and other quantum algorithms
that have collective properties. 
The library is adaptable to simulate
collective phenomena 
in various physical platforms, 
such as trapped ions
\cite{Schindler_2013}, 
ultracold atoms in optical lattices
\cite{PhysRevLett.129.090403},
Rydberg atom arrays in optical tweezers
\cite{PhysRevLett.123.260505},
and nitrogen-vacancy centers
\cite{PhysRevLett.110.156402}.

This paper is organized as follows. 
Section \ref{sec2} introduces quantum computing with
collective processes in ensembles of two-level systems.
Section~\ref{sec3} details the structure of \texttt{tqix.pis} library and its benchmark results are given in Section~\ref{sec4}. 
Section~\ref{sec5} is devoted to 
applications on spin squeezing 
and quantum phase transition. 
A brief conclusion is given in Section~\ref{sec6}.
The full code for generating 
figures in the paper can be 
found in the Appendices.

\section{Quantum computing with 
collective processes in ensembles of two-level systems}
\label{sec2}
\subsection{Collective processes in 
ensembles of two-level systems}
Consider an ensemble of $N$ two-level particles
characterized by the
single and collective angular momentum operators
$J_{\alpha}^{(n)} = \frac{1}{2}\sigma_{\alpha}^{(n)}$
and 
$J_{\alpha} = \sum_n J_\alpha^{(n)}$,
respectively, where
$\sigma_\alpha$, 
$\alpha = \{x, y, z\}$,
are Pauli matrices.
The joint Hilbert space of the ensemble 
is a composite $\mathscr{H}_{\rm E} = \mathscr{H}^{(1)}
\otimes\cdots\otimes\mathscr{H}^{(N)}$
with ${\rm dim}(\mathscr{H}_{\rm E}) = 2^N$.
A generic mixed state 
is given by
a $(2^N \times 2^N)$-matrix as
\begin{align}\label{eq:rho}
\notag    \rho = &\sum_{\substack{
            m_1, m_2,\cdots,m_N\\
            m_1', m_2',\cdots,m_N'}}
     \rho_{m_1, m_2,\cdots,m_N;
            m_1', m_2',\cdots,m_N'}\\
    & \times\
    |m_1, m_2,\cdots,m_N\rangle
    \langle m_1', m_2',\cdots,m_N'|,
\end{align}
where the product basis is
$|m_1,m_2,\cdots,m_N\rangle=
|m_1\rangle\otimes|m_2\rangle\otimes\cdots\otimes
|m_N\rangle$,
with $m_n=\pm\frac{1}{2}$
are eigenvalues of $J_z^{(n)}$.

Under the collective processes 
\cite{PhysRevA.78.052101,PhysRevA.81.032104}, 
the product basis is represented 
by an irreducible representation 
(irrep) basis $\{|j,m,i\rangle\}$,
where $j\le N/2$ is the total angular momentum with
$j_{\rm min} = \mod(N/2)/2,\ j_{\rm max}
= N/2$, and $|m| \le j$.
The quantum number $i = 1, 
\cdots, d_N^j$ 
distinguishes $d_N^j$ 
degenerate irreps, where
\begin{align}\label{eq:dNj}
d_N^j = \dfrac{N!(2j+1)}
{(N/2-j)!(N/2+j+1)!}
\end{align}
is the number of 
ways to combine $N$ particles 
that gets the total angular momentum $j$
\cite{Mikhailov_1978}.
%
The quantum state is recast 
in the irrep basis as 
\begin{align}\label{eq:mix-state-in-irr}
    \rho = 
    \sum_{\substack{
            j,m,i\\
            j',m',i'}}
    \rho_{jmi,j'm'i'}|j,m,i\rangle
    \langle j',m',i'|.
\end{align}

The state $\rho$
is permutation invariant (PI) 
if it does not change
under a permutation operator 
$P_\pi$ of a permutation $\pi$, i.e.,
$P_\pi\rho P_\pi^\dagger = \rho\ \forall \pi$
\cite{PhysRevA.88.012305}.
Decompose $P_\pi$
into the multiplicity subspaces 
$\mathscr{H}_j\otimes\mathscr{K}_j$ 
following the Schur-Weyl duality
\cite{PhysRevA.93.012320} 
as
$
    P_\pi  
    = \oplus_{j = j_{\rm min}}^{j_{\rm max}}
    \bm I_{\mathscr{H}_j}\otimes P_j(\pi),
$    
where $ 
\oplus_{j = j_{\rm min}}^{j_{\rm max}}
\mathscr{H}_j\otimes\mathscr{K}_j = \mathscr{H}_{\rm E}$,
${\rm dim}(\mathscr{H}_j) = 2j + 1$
and ${\rm dim}(\mathscr{K}_j) = d_N^j$,
that $P_\pi$ only acts on 
the irrep subspace $\mathscr{K}_j$,
which causes a general permutation symmetry, 
i.e., $\rho_{jmi,jm'i} = \rho_{jmi,jm'i'}
\ \forall i,i' \in [1, d_N^j]$,
then a permutation-invariant state is given by
\begin{align}
    \rho_{\rm PI} = \bigoplus_{j = j_{\rm min}}^{j_{\rm max}}
    \rho_j\otimes\bm I_{\mathscr{K}_j}
    = \sum_{j,m,m'}
    \rho_{jmm'}|j,m\rangle
    \langle j,m'|
    \otimes\bm I_{\mathscr{K}_j}.
\end{align}
Hence, we can ignore the identical irrep label $i$ 
and recast the state in a collective form as
\cite{PhysRevA.78.052101}
\begin{align}\label{eq:rho-in-dicke}
    \rho_{\rm C} =
    \bigoplus_{j = j_{\rm min}}^{j_{\rm max}}
    \rho_j = \sum_{j,m,m'}
    \rho_{jmm'}|j,m\rangle
    \langle j,m'|,
\end{align}
where the effective amplitude 
$\rho_{jmm'}$
and the effective density matrix elements
$|j,m\rangle\langle j,m'|$ are given by
\begin{align}\label{eq:dickebas}
    \rho_{jmm'}|j,m\rangle
    \langle j,m'| = 
    \frac{1}{d^j_N}
    \sum_{i=1}^{d^j_N}|j,m,i\rangle
    \langle j,m',i|.
\end{align}
The effective basis $\{|j,m\rangle\}$
is known as the Dicke basis \cite{PhysRev.93.99},
i.e., $
    |j,m\rangle = 
    \frac{1}{\sqrt{d^j_N}}
    \sum_{i=1}^{d^j_N}|j,m,i\rangle,
$
that is the eigenstates of the 
collective angular momentum operators
\begin{align}
\bm{J}^2|j,m\rangle &= j(j+1)|j,m\rangle, \label{eq:j2}\\ 
J_z|j,m\rangle &= m|j,m\rangle, \label{eq:jz}\\ 
J_\pm|j,m\rangle &= \sqrt{(j\mp m)(j\pm m+1)}
|j,m\pm1\rangle, \label{eq:jpm}
\end{align}
where 
$\bm J^2 = J_x^2 + J_y^2 + J_z^2$ and 
$J_\pm = J_x\pm iJ_y.$
The collective angular momentum operators 
obey the commuting relation
\begin{align}\label{eq:uncertainty relation commuting}
     [J_\alpha,J_\beta] = i\epsilon_{\alpha\beta\gamma}J_\gamma,
\end{align}
where $\{\alpha,\beta,\gamma\} 
\in \{x,y,z\}$, 
and $\epsilon_{\alpha\beta\gamma}$ 
is the Levi-Civita symbol.

Under the collective processes, 
the Hilbert space $\mathscr{H}_N$ 
reduces to a collective subspace 
$\mathscr{H}_{\rm C} \sim O(N^2) \subset \mathscr{H}_{\rm E} $ as 
\cite{PhysRevA.78.052101,PhysRevA.81.032104}
\begin{align}\label{eq:hc-dim}
\notag  {\rm dim} (\mathscr{H}_{\rm C}) 
    & = \bigoplus_j {\rm dim}(\mathscr{H}_j) \\
    & = \begin{cases}
    (N+3)(N+1)/4, & \text{for odd $N$},\\
    (N+2)^2/4, & \text{for even $N$}.
  \end{cases}
\end{align}

\begin{figure}[t]
\centering
\includegraphics[width=9cm]{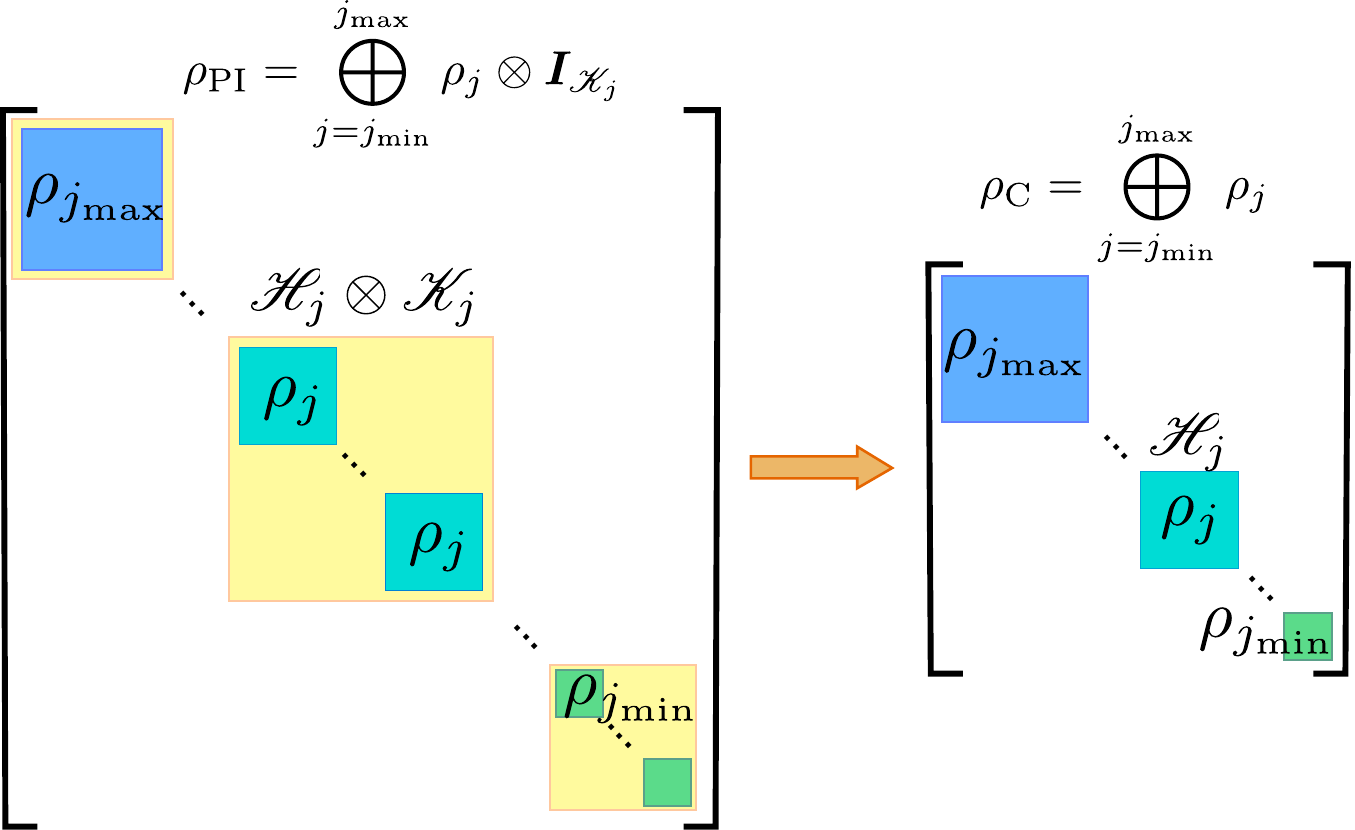}
\caption{
Block-diagonal structure of a quantum state in 
the irrep representation with permutation invariant
(left) and its corresponding in the collective 
representation (right).
Each yellow block represents the 
subspace $\mathscr{H}_j\otimes\mathscr{K}_j$ 
and contains $d_N^j$ 
identical density matrices $\rho_j$.
In the collective form, we ignore the
identity of density matrices by removing
the subspace $\mathscr{K}_j$.
}
\label{fig:1}
\end{figure}

Notable that both the PI and 
collective states 
have the $j$-irrep block-diagonal structure 
where all degeneracy with the same $j$
are presented in the same block.
See figure~\ref{fig:1} 
for the block-diagonal structure 
of these quantum states.
Similarly, the collective angular 
momentum operators $\{J_\alpha\}$
are represented by 
block-diagonal of spin-$j$ operators
$\{S_\alpha^{(j)}\;, \forall j 
\in [j_{\rm min}, j_{\rm max}]\}$.




Especially in the symmetry group 
\cite{PhysRevA.67.022112,Duan2001,PhysRevA.66.023818}, 
a symmetry mixed state is
a particular case of 
the collective state with 
$\rho_{jmm'} = 0$ for $j\ne N/2$
and is presented in 
the first $j$-irrep block 
$\mathscr{H}_{{N}/{2}}$ with 
the dimension dim$(\mathscr{H}_{{N}/{2}}) = N+1$.
The symmetry states characterize 
the ensemble phenomena of multiparticle, such as 
entanglement
\cite{PhysRevA.67.022112,PhysRevA.88.012305}, 
spin-squeezing
\cite{PhysRevA.47.5138,PhysRevLett.83.1319},
and zero-temperature quantum phase transitions
\cite{PhysRevLett.100.040403,
PhysRevLett.90.044101,PhysRevA.7.831,
CARMICHAEL197347,PhysRevE.67.066203}.

\subsection{Collective quantum states}
An important collective quantum state
is the ground state where all spins
are down, 
i.e., $|g\rangle^{\otimes N}
=|\downarrow\cdots\downarrow\rangle$.
In the Dicke basis, it gives
$|g\rangle^{\otimes N} = \big|\frac{N}{2}, 
- \frac{N}{2}\Big\rangle$
and its density matrix locates 
in the bottom-right position 
of the first irrep block.
Another one is the excited state,
$|e\rangle^{\otimes N} =
|\uparrow\cdots\uparrow\rangle$,
which is given by 
$|e\rangle^{\otimes N} = 
\big|\frac{N}{2}, \frac{N}{2}\Big\rangle$,
and the density matrix locates 
on the top-left of the first irrep block.
An entangle GHZ state is a superposition 
of the ground and excited states
as 
\begin{align}\label{eq:ghz}
    |{\rm GHZ}\rangle =
    \dfrac{1}{\sqrt{2}}
    \Bigg(
    \big|\textstyle\frac{N}{2},\frac{N}{2}\Big\rangle
    +\big|\frac{N}{2},-\frac{N}{2}\Big\rangle
    \Bigg).
\end{align}
Besides, a literature well-known coherent spin state 
(CSS) is a product state that resembles 
a coherent state 
of a classical harmonic oscillator
\cite{Radcliffe_1971,PhysRevA.6.2211}.
It explicitly expands 
in terms of a linear combination of 
$|\frac{N}{2}, m\rangle$ elements
as
\begin{widetext}
\begin{align}\label{eq:css}
    |\theta,\phi\rangle_{\rm CSS}=
    \sum_{m=-\frac{N}{2}}^{\frac{N}{2}}
    \sqrt{\binom{N}{\frac{N}{2}+m}}
    \Big(\cos\textstyle\frac{\theta}{2}\Big)^{\frac{N}{2}+m}
    \Big(\sin\textstyle\frac{\theta}{2}
    e^{-i\phi}\Big)^{\frac{N}{2}-m}
    \Big|\frac{N}{2},m\Big\rangle,
\end{align}
\end{widetext}
where $0\le\theta\le\pi$ is the polar angle,
and $0\le\phi\le 2\pi$ denotes the azimuth angle
in the spherical coordinates.

We emphasize that all these states 
obey the symmetry and are thus represented 
in the subspace $\mathscr{H}_{N/2}$.
However, under noise,
such as decoherence, 
symmetry states decay from the 
maximum $j_{\rm max}$-irrep block 
and transfer 
to the lower 
blocks, 
resulting in the
full collective subspace $\mathscr{H}_{\rm C}$.
(See Ref.~\cite{PhysRevA.81.032104}
for an example.)

\subsection{Collective operators}
Collective operators are 
unitary transformations
over all particles of the ensemble.
We first introduce the rotation operator
denoted by $R_{\bm n}(\theta, \phi)$.
It rotates the original 
state counterclockwise 
by a polar angle $\theta$ 
around the $xy$ 
plane and then by 
an angle $\phi$ 
around the $z$ axis
\cite{MA201189}
\begin{align}\label{eq:rn}
    R_{\bm n}(\theta,\phi) 
    = e^{-i\theta \bm J_{\bm n}}
    = {\rm exp}
    [i\theta (J_x\sin\phi - J_y\cos\phi)],
\end{align}
where ${\bm J} = (J_x, J_y, J_z),\
{\bm n} = (-\sin\phi, \cos\phi, 0)$,
and ${\bm J}_{\bm n} = {\bm J}\cdot {\bm n}$.
For example, a CSS state 
can be prepared from the rotation 
$R_{\bm n}(-\theta,\phi)$ of the
ground state,
i.e., $|\theta,\phi\rangle_{\rm CSS}
= R_{\bm n}(-\theta,\phi)|
\textstyle\frac{N}{2},-\frac{N}{2}\rangle$
\cite{PhysRevA.6.2211}; 
or an excited state is given by a rotation
$R_{\bm n}(\pi,\phi)|\frac{N}{2},-\frac{N}{2}\rangle$.
Specific cases of the rotation operator are rotation around 
the $\alpha$ axis, $\alpha \in \{x,y,z\}$ as
\begin{align}\label{eq:Ra}
    R_\alpha(\theta) = 
    e^{-i\theta J_\alpha}.
\end{align}
These rotations allow us
to transform quantum states everywhere
in the spherical representation but not deform the states.

Hereafter, let us introduce other 
classes of transformation based on 
the nonlinear interactions
that can deform the quantum states.
They include one-axis twisting (OAT)
\cite{PhysRevA.47.5138,
PhysRevLett.86.4431,
Sorensen2001,
PhysRevA.65.043610,
Gross2010,Riedel2010,PhysRevA.66.022314,
Schulte2020ramsey}, 
two-axis countertwisting (TAT)
\cite{PhysRevA.47.5138,Borregaard_2017,
PhysRevA.96.013823,PhysRevA.99.043840,
PhysRevLett.107.013601},
twist-and-turn (TNT)
\cite{PhysRevA.67.013607,
doi:10.1126/science.1250147,
PhysRevA.92.023603,PhysRevA.63.055601},
and an important global 
M\o{}lmer-S\o{}rensen
opterator (GMS)
\cite{PhysRevLett.82.1835,
PhysRevLett.82.1971,
Maslov_2018,Groenland_2020,
van_de_Wetering_2021}.

An OAT transformation along 
the $\alpha$ axis is defined by
\begin{align}\label{eq:oat}
    U_{\rm OAT} = e^{-it\chi J_\alpha^2},
\end{align}
where through the paper, we use 
$\hbar = 1$, 
$\chi$ is the nonlinearity 
interparticle interaction,
and $t$ is the time.
It leads to a rotation proportional to 
$J_\alpha,\ \alpha\in\{x,y,z\}$ 
and twist 
the quantum fluctuations 
of the ensemble.
Similarly, a TAT transformation is defined by
\begin{align}\label{eq:tat}
    U_{\rm TAT} = e^{-it\chi (J_\alpha^2 - J_\beta^2)},
\end{align}
where $\{\alpha,\beta\} \in 
\{x,y,z,+,-\}$.
Next, for the TNT transformation,
we have 
\begin{align}\label{eq:tnt}
    U_{\rm TNT} = e^{-it
    (\chi J_\alpha^2-\Omega J_\beta)},
\end{align}
where $\Omega$ is 
the linearly coupling strength.
Set $\Lambda = N\chi/\Omega$,
then for $\chi\ll\Omega$,
the TNT transformation
reduces to the OAT one.
Finally, the definition of 
a global M\o{}lmer-S\o{}rensen transformation 
is an entangling gate that operates on all particles
of the ensemble \cite{M_ller_2011}
\begin{align}\label{eq:MS}
    U_{\rm GMS} = 
    e^{-i\theta (J_x\cos\phi  
    + J_y\sin\phi)^2}.
\end{align}
It encompasses the pairwise 
and is given by two angles $\theta$
and $\phi$. 
For $\theta = \pi/2$, the GMS gate is
maximally entangling, i.e., GHZ state
\cite{PhysRevLett.82.1835}.

These transformations deform 
the collective states and 
result in the entangled or squeezed 
effect in the ensemble systems
and thus exhibit various applications
from quantum-enhanced meteorology
\cite{Hosten2016,
PhysRevA.97.032116,
PhysRevLett.124.060402,
PhysRevResearch.4.013236}
to quantum sensors, and atomic clocks
\cite{PRXQuantum.3.020308,PhysRevX.11.041045}.
Besides, other transformations
for creating highly quantum-enhanced states
include light-to-atoms transformation
\cite{PhysRevLett.79.4782,
PhysRevA.60.1491,
PhysRevLett.88.070404,
PhysRevLett.96.133601,
PhysRevA.90.063630,
PhysRevA.91.041802},
quantum nondemolition measurement (QND)
\cite{doi:10.1126/science.209.4456.547,
Hosten2016,PhysRevLett.85.1594,
Louchet_Chauvet_2010,
Yang2020,RevModPhys.68.1,
Grangier1998},
and adiabatically quantum phase transition 
\cite{PhysRevLett.111.180401,
doi:10.1126/science.aag1106,
PhysRevA.97.032339},
to name a few.

\subsection{Depolarizing noise in the collective systems}
\label{sub24}
Depolarizing noise is a typical type of quantum noise
\cite{nielsen_chuang_2010}. 
It can be defined through bit-flip,
phase-flip, and bit-phase-flip channels 
for every single particle in the circuit
\cite{7927034}. 
Let $\rho = \sum_{j,m,m'} \rho_{jmm'}
|j,m\rangle\langle j,m'|$ is the initial quantum state. 
Through a depolarizing channel applied to 
all particles in the ensemble,
the state transforms to 
\cite{nielsen_chuang_2010}
\begin{align}\label{eq:dep_noise}
\notag \rho_1 = (1-\epsilon)\rho 
+ & \sum_{n=1}^N 
\Big(
\epsilon_x J_x^{(n)}\rho J_x^{(n)} \\
&+ \epsilon_y J_y^{(n)}\rho J_y^{(n)}
+ \epsilon_z J_z^{(n)}\rho J_z^{(n)}
\Big),
\end{align}
where $0\le \epsilon \le 1$ 
is the depolarized probability, 
and $\epsilon_{i}, \ i\in\{x,z,y\}$ are  
the probabilities the of 
bit-flip,
phase-flip, 
and bit-phase-flip errors, respectively. 
In this paper, we restrict to symmetry noise channels,
i.e., $\epsilon_x = 
\epsilon_y =\epsilon_z = \epsilon'$.
Here, $\epsilon'$ obeys the normalized condition
that ${\rm tr}(\rho_1) = 1$, or
$\epsilon' = \epsilon/{\rm tr}\Big[\sum_{n=1}^N
\big(
J_x^{(n)}\rho J_x^{(n)}
+J_y^{(n)}\rho J_y^{(n)}
+J_z^{(n)}\rho J_z^{(n)}\big)
\Big]
$.
Substituting 
\begin{align}\label{eq:jpm1}
    J_x^{(n)} = \dfrac{1}{2}
    \Big(J_+^{(n)} + J_-^{(n)}\Big);\
    J_y^{(n)} = \dfrac{-i}{2}
    \Big(J_+^{(n)} - J_-^{(n)}\Big)
\end{align}
into Eq.~\eqref{eq:dep_noise}
and set 
\begin{align}\label{eq:dep_noise_rho_prime}
\rho' = 
\sum_{n=1}^N
\Big[
\frac{1}{2}\Big(
J_+^{(n)}\rho J_-^{(n)}
+J_-^{(n)}\rho J_+^{(n)}
\Big)
+J_z^{(n)}\rho J_z^{(n)}
\Big],
\end{align}
we rewrite Eq.~\eqref{eq:dep_noise} as

\begin{align}\label{eq:dep_noise_rewrite}
\rho_1 = (1-\epsilon)\rho + \epsilon\dfrac{\rho'}{{\rm} tr(\rho')}.
\end{align}

Following Refs.
\cite{PhysRevA.78.052101,
PhysRevA.81.032104,
PhysRevA.98.063815,
PhysRevA.102.022602}, 
we derive the terms
$
    \sum_{n=1}^N J_k^{(n)}
    |j,m\rangle\langle j,m'|
    J_l^{(n)\dagger} 
$ for all $\{k,l\}\in 
\{+,-,z\}$,
and compute $\rho_1$ straightforwardly. 

Under the noise, the permutation symmetry in 
the system is broken, 
resulting in a deviation from 
$|j,m\rangle$ to $|j\pm 1,m\pm 1\rangle$.
Thus, the final state spans in 
the collective Hilbert space 
$\mathscr{H}_{\rm C}$
with all $j$-irrep blocks for all 
$j_{\rm min}\le j\le j_{\rm max}$.

In Sec.~\ref{sec3},
we introduce \texttt{tqix.pis} 
library 
for 
quantum dynamics simulation of
ensemble systems 
in the Dicke basis
based on the theoretical framework 
described here.

\section{
Quantum simulation with 
collective processes 
in ensembles of two-level systems}
\label{sec3}

The simulation is based on \texttt{tqix}
package, a quantum toolbox
for quantum measurement, metrology,
and tomography \cite{HO2021107902}. 
Here, we develop \texttt{tqix.pis} library for 
simulating 
various quantum dynamics of
spin ensembles in Dicke basis.

\subsection{Features of the library}

\begin{table*}
\small
\centering
\caption {List of basic quantum gates (function type)
built in Gate class of the 
library \texttt{tqix.pis}.
In each gate, the additional argument
*args (float) allows to add the depolarized probability (noise parameter).
} \label{tab:1}
\setlength{\tabcolsep}{6pt}
\begin{tabular}{m{2.cm} m{6.5cm} m{7.cm}}
 \hline
  Name & Syntax & Description\\
  \hline\hline
$R_x$ gate 
  & \texttt{\textcolor{cyan}{RX}($\theta$,*args)}
  & Rotate the quantum state at an angle $\theta$
  around the $x$ axis, i.e., 
  ${\rm RX}(\theta) = e^{-i\theta J_x}$ \\
$R_y$ gate 
  & \texttt{\textcolor{cyan}{RY}($\theta$,*args)}
  & Rotate the quantum state at an angle $\theta$
  around the $y$ axis, 
  i.e., ${\rm RY}(\theta) = e^{-i\theta J_y}$ \\
$R_z$ gate 
  & \texttt{\textcolor{cyan}{RZ}($\theta$,*args,)}
  & Rotate the quantum state at an angle $\theta$
  around the $z$ axis, i.e., 
  ${\rm RZ}(\theta) = e^{-i\theta J_z}$ \\
$R_{\bm n}$ gate 
  & \texttt{\textcolor{cyan}{RN}($\theta,\phi$,*args)}
  & ${\rm RN}(\theta,\phi) = e^{-i\theta(J_x\sin\phi
  -J_y\cos\phi)}$ \\
$R_{\bm +}$ gate 
  & \texttt{\textcolor{cyan}{R\_plus}($\theta$,*args)}
  & ${\rm R\_plus}(\theta) = e^{-i\theta J_+}$ \\
$R_{\bm -}$ gate 
  & \texttt{\textcolor{cyan}{R\_minus}($\theta$,*args)}
  & ${\rm R\_minus}(\theta) = e^{-i\theta J_-}$ \\  
$R_x^2$ gate 
  & \texttt{\textcolor{cyan}{RX2}($\theta$,*args)}
  & ${\rm RX2}(\theta) = e^{-i\theta J_x^2}$ \\
$R_y^2$ gate 
  & \texttt{\textcolor{cyan}{RY2}($\theta$,*args)}
  & ${\rm RY2}(\theta) = e^{-i\theta J_y^2}$ \\
$R_z^2$ gate 
  & \texttt{\textcolor{cyan}{RZ2}($\theta$,*args)}
  & ${\rm RZ2}(\theta) = e^{-i\theta J_z^2}$ \\
  \hline
OAT gate
  & \texttt{\textcolor{cyan}{OAT}($\theta$,gate\_type,*args)}
  & One-axis twisting gate, e.g.,
  OAT($\theta$,\squote \texttt{x}\squote) 
  = $e^{-i\theta J^2_{x}}$\\
  & \texttt{------------} & \\
  & $\theta = t\chi$: rotation angle & \\
  & \texttt{gate\_type} =  $\{$\squote \texttt{x}\squote, 
                               \squote \texttt{y}\squote,
                               \squote \texttt{z}\squote $\}$ &\\
  \hline
TAT gate
  & \texttt{\textcolor{cyan}{TAT}($\theta$,gate\_type,*args)}
  & Two-axis twisting gate, e.g.,
  TAT($\theta$,\squote$\alpha\beta$\squote) = $e^{-i\theta (J_\alpha^2-J_\beta^2)}$\\
  & \texttt{------------} & \\
  & \texttt{gate\_type} =  \squote$\alpha\beta$\squote, $\forall$ 
  $\{\alpha,\beta\} \in \{x, y, z, plus, minus\}$ &\\
  \hline
TNT gate
  & \texttt{\textcolor{cyan}{TNT}($\theta, \Lambda$,gate\_type,*args)}
  & Twist and Turn gate, e.g.,
  TNT($\theta, \Lambda$,\squote$\alpha\beta$\squote) 
  = $e^{-i\theta (J_\alpha^2-\frac{N}{\Lambda}J_\beta)}$\\
  & \texttt{------------} & \\
  & \texttt{gate\_type} =  \squote$\alpha\beta$\squote, $\forall$ 
  $\{\alpha,\beta\} \in \{x, y, z, plus, minus\}$ &\\
  \hline
GMS gates 
  & \texttt{\textcolor{cyan}{GMS}($\theta,\phi$,*args)}
  & Global M\o{}lmer-S\o{}rensen gate, e.g.,
  GMS($\theta, \phi$) = $e^{-i\theta (J_x\cos\phi  
    + J_y\sin\phi)^2}$\\
  \hline
\end{tabular}
\end{table*}
\begin{itemize}
    \item Assist a large number of particles with the collective processes of the ensemble.
    \item Assist fast simulation with parallelizing multi-core processors and Graphics Processing Units (GPUs).
\end{itemize}

\subsection{Structure of the program}
A basic quantum circuit
includes a quantum register (quantum state), 
quantum gates, and measurements.
In \texttt{tqix.pis}, 
the computational basis is 
Dicke basis $|j,m\rangle\langle j,m|$,
represented by a sparse matrix. 
The quantum register is 
a density matrix in the Dicke basis,
initially prepared in the ground state,
i.e., $\rho_0 = \textstyle
|\frac{N}{2},-\frac{N}{2}\rangle
\langle\frac{N}{2},-\frac{N}{2}|$.
This is a symmetry collective state
spans in the subspace $\mathscr{H}_{N/2}$,
and thus it has the dimension of 
$(N+1)\times(N+1) \ll 2^N\times 2^N$.
This is a highlighted feature of the library
that allows us to compute 
a large number of particles.
Similarly, quantum gates are 
collective operators,
e.g., $K$,
that apply to the quantum register and 
transform it into the evolved state
\begin{align}\label{eq:evoled_state}
    \rho = K\rho_0 K^\dagger.
\end{align}
A list of build-in quantum gates and their syntax
are given in Tab.~\ref{tab:1}. 
These gates are represented 
by sparse matrices with the
same dimension as the quantum register.
Finally, the measurement 
on the Dicke basis gives the probability
\begin{align}\label{eq:prob}
    P(j,m) = \tr \big[\rho\cdot |j,m\rangle\langle j,m|\big]
    = \langle j, m|\rho|j,m\rangle.
\end{align}
To mimic the experiential probability,
we apply the \texttt{cdf} 
back-end in \texttt{tqix}
\cite{HO2021107902}.

The following example code
shows how to execute a simple 
quantum circuit with 
a rotation gate $R_{\bm n}$ using 
the \texttt{circuit, RN,} and 
\texttt{measure} functions
in the library \texttt{tqix.pis}:

\begin{figure}[t]
\centering
\includegraphics[width=8.6cm]{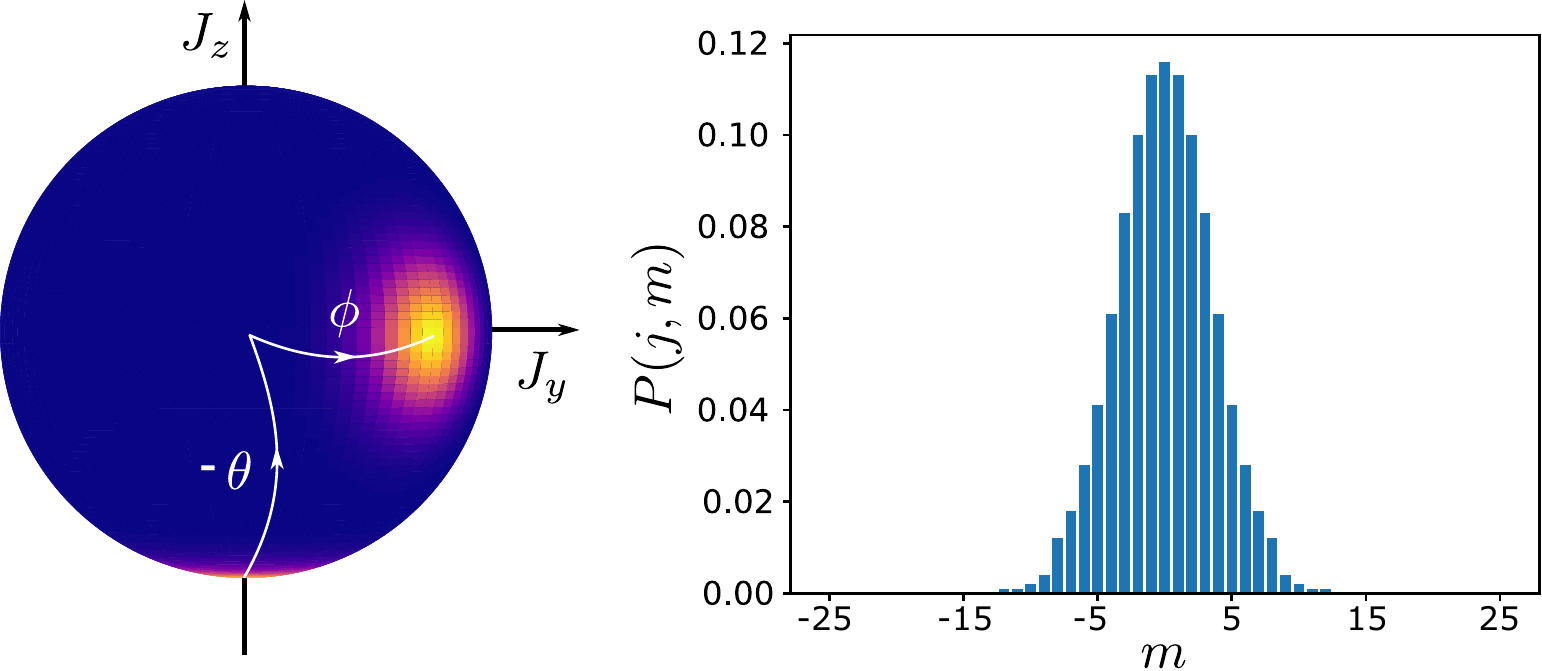}
\caption{
(Left) Husimi visualization for the circuit 
initially prepared in the ground state (the South pole)
and then transferred into a CSS state by applying
a rotation gate ${\rm RN}(-\pi/2,\pi/4)$.
(Right) Probability distribution $P(j,m)$ of the 
final state measuring on the Dicke basis $|j,m\rangle$.
Plot for \#particles $N = 50$.
}
\label{fig:2}
\end{figure} 

\begin{lstlisting}
from tqix import *
from tqix.pis import *
import numpy as np

N = 50 #particles
qc = circuit(N) #create circuit
qc.RN(-np.pi/2,np.pi/4) #apply RN
prob = qc.measure(num_shots=1000) #measure

#to get state information
psi = qc.state #sparse matrix
psi = qc.state.toarray() #full matrix
\end{lstlisting}

The results are given in figure~\ref{fig:2}.
The left figure is the Husimi visualization
for the circuit's state 
initially prepared in the ground state 
(the South pole). Under the rotation gate
$\rm {RN}(-\pi/2,\pi/4)$, 
the state rotates around the $y$ 
axis at an angle $\theta = -\pi/2$
and then rotates around the $z$ axis 
at an angle $\phi = \pi/4$.
The final state becomes a CSS state,
where its probabilities 
measured in the Dicke basis
are exhibited in the right figure.
(See the full code in \ref{appA}.)

To compute the expectation value of
a given observable, such as $J_z$,
we execute the 
\texttt{expval} function as
\begin{lstlisting}
expJz = qc.expval('Jz')
\end{lstlisting}
We emphasize that the observable
must obey the collective process,
such as the collective angular momentum operators
$\{J_\alpha\}$, for $\alpha\in \{x,y,z,+,-\}$.
The program supports 
these observables $\{J_\alpha\}$ include
\texttt{\squote Jx\squote, 
\squote Jy\squote,
\squote Jz\squote, 
\squote J\_plus\squote, 
\squote J\_minus\squote} 
and $\{J_\alpha^2\}$ include 
\texttt{\squote Jx2\squote, 
\squote Jy2\squote, 
\squote Jz2\squote, 
\squote J\_plus2\squote, 
\squote J\_minus2\squote}.
For an arbitrary self-defined observable, such as
a linear combination of these listed observables,
the users can compute manually 
from the quantum state and 
the self-defined observable.
For example, 
the following script computes 
the expectation value 
of $J_x + J_y + J_z$
\begin{lstlisting}
Jx = qc.Jx()
Jy = qc.Jy()
Jz = qc.Jz()
Js = Jx + Jy + Jz
psi = qc.state
expectJ = np.trace(psi@Js)
\end{lstlisting}

Finally, when we apply a quantum gate onto the circuit, 
it also causes noises due to the assumption 
of the imperfection of the gate.
In the current version of the library, 
we restrict to the typical depolarizing noise as described  
in subsection \ref{sub24}.
To include the noise, 
we use \texttt{noise} option in 
the quantum gate function, which is the 
depolarized probability 
(or noise parameter) $\epsilon$.
For example, we add the noise with $\epsilon = 0.01$
into the \texttt{RN} gate as following
\begin{lstlisting}
qc.RN(-np.pi/2,np.pi/4, noise = 0.01)
\end{lstlisting}

\subsection{Additional options for running multi-core processors
and GPUs}
By default, the program executes 
on CPUs with one processor.
However, it also allows for running 
parallel multi-core processors, 
which assists the running time.
Technically, we use the python multiprocessing 
module to create multiple processors 
for handling different parts of the input list.
To execute the program in multi-core processors, 
we add the option \texttt{num\_process} into 
the \texttt{circuit}. This is an integer number
stands for the number of processors. 
For example, one can create a quantum register
with 50 particles and run it with 25 processors 
by using the following command
\begin{lstlisting}
qc = circuit(N = 50, num_process = 25)
\end{lstlisting}

Similarly, the library supports 
running on GPU devices, 
which can outperform the CPUs
in some cases. 
To execute the program with GPUs,
we add the option \texttt{use\_gpu} 
into the \texttt{circuit}.
For example, 
\begin{lstlisting}
qc = circuit(N = 50, use_gpu = True)
\end{lstlisting}
Note that to run with GPUs, 
the users must turn off the option 
\texttt{num\_process} or leave it as one.
In GPUs, the sparse-matrix structure 
will convert into the tensor structure 
using \texttt{pyTorch} library.

%

%
\begin{figure}[t]
\centering
\includegraphics[width=8.6cm]{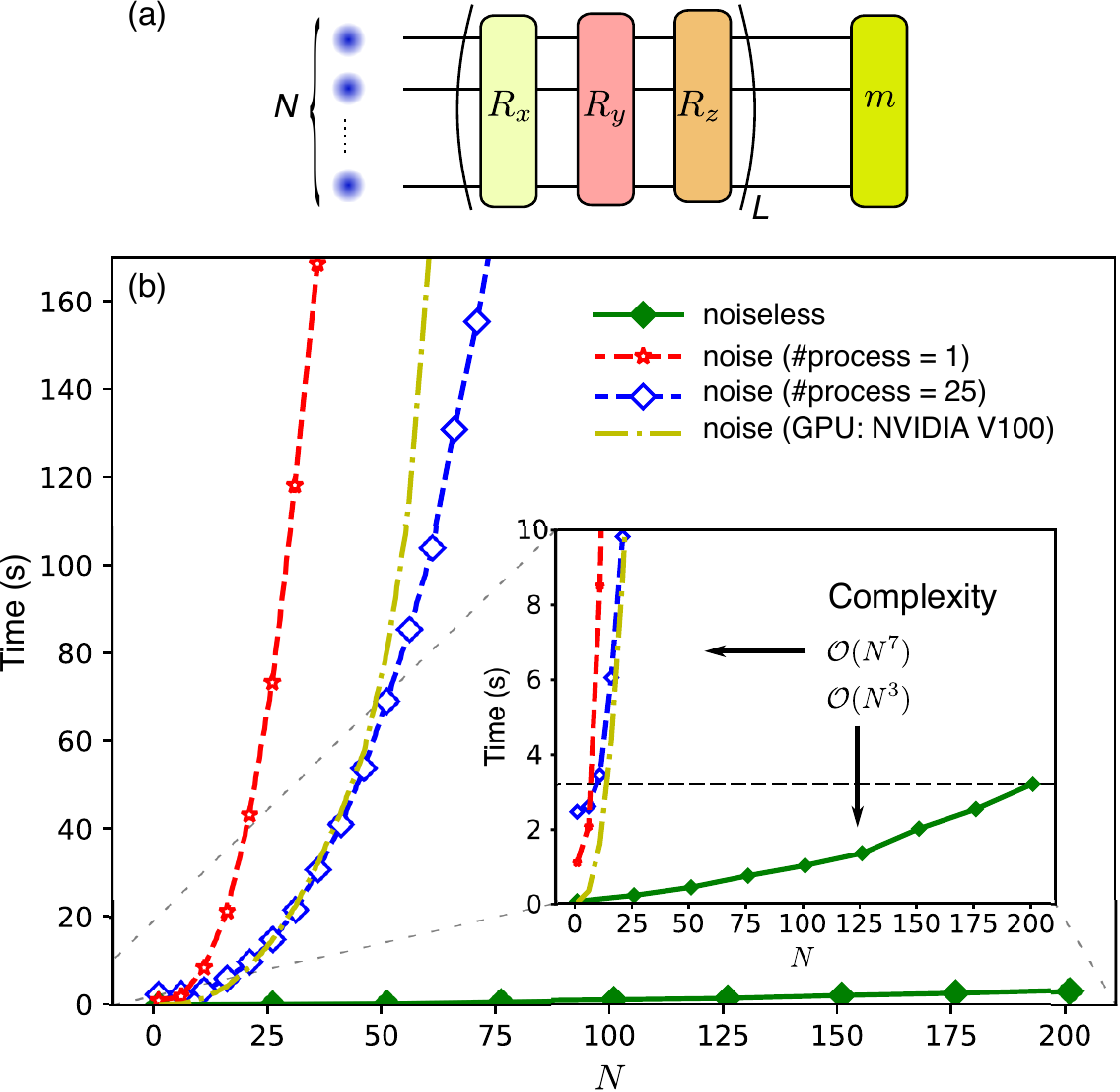}
\caption{
(a) Benchmarking quantum circuit. 
The circuit consists of $N$ particles,
transfers under $L$ 
layers of rotation gates 
$R_x, R_y, R_z$, 
and measures in the Dicke basis. 
(b) The running time 
versus the number of particles $N$ for different
cases of with/without noise and CPUs/GPUs. 
Inset: zoom in the $y$ 
axis from 0 to 10 seconds.
The results are shown for $L=3$.
}
\label{fig:3}
\end{figure} 

\section{Benchmark results}
\label{sec4}
For benchmarking the library,
we consider the execution time
versus the number of particles $N$.
The benchmarking circuit 
is shown in Fig.~\ref{fig:3}(a),
which consists of 
a set of rotation gates 
$\{R_x, R_y,$ and $R_z\}$ with $L$ times 
and the measurement of the final state.
The running time versus $N$ for
$L=3$ is shown in Fig.~\ref{fig:3}(b).
For the noiseless case
(filled diamond), 
the circuit executes up to $N=200$, 
consuming around 3 seconds 
with 1 processor CPU
(see the inset Fig.~\ref{fig:3}(b)).
The complexity is 
$\mathcal{O}(N^3)$.
Under the presence of noise, 
the running time rapidly increases with $N$.
The complexity is 
$\mathcal{O}(N^7)$ for CPU regardless 
of the number of processors.
Nevertheless, by
using multi-processing on CPUs, 
we can reduce the running time
of adding noise by $p$ times, 
with $p$ being the computer's 
maximum processors.
Hence, the running time 
for 25 processors (open diamond)
is faster than the 1 processor (open star).

We next evaluate the performance of the GPU (NVIDIA V100) (dashed-dot curve)
and compare it with the 25-processor CPU. 
The complexity performing on the GPU is $\mathcal{O}(N^7)$.
For $N < 25$, it offers a better running time,
while increasing $N$ will slow down the running time.
This result relies on 
the different matrix structures 
in these two devices,
i.e., the CPU uses sparse matrices, 
whereas the GPU uses tensor matrices.
For small $N$, the tensor dimension is small,
the GPU thus displays its advantage overcome the CPU,
and for larger $N$, the tensor dimension increases
while the sparse matrices are unaffected.
As a result, the performance of CPU
gradually overcomes the GPU.
The full code for producing
Fig.~\ref{fig:3} is given in 
\ref{appB}.

The time complexities 
of a quantum circuit 
with various cases 
are listed in Tab.~\ref{tab:2}.
We evaluate the complexity by
analyzing the number 
of operations performed on
the code’s statements.
Details are given in \ref{appC}.

\begin{table}[!ht]
\small
\centering
  \caption{The complexities 
  of a quantum circuit 
  for different cases 
  as shown in the list.}
  \begin{tabular}{|K{2.5cm}| K{2.5cm}| K{2.5cm}|}
    \hline
     & Noiseless & Noisy  \\
    \hline
    Symmetry state & $\mathcal{O}(N^3)$ & $\mathcal{O}(N^6)$  \\
    \hline
     Collective state & $\mathcal{O}(N^6)$ & $\mathcal{O}(N^7)$ \\
    \hline
  \end{tabular}
  \label{tab:2}
\end{table}

\begin{figure}[t]
\centering
\includegraphics[width=8.6cm]{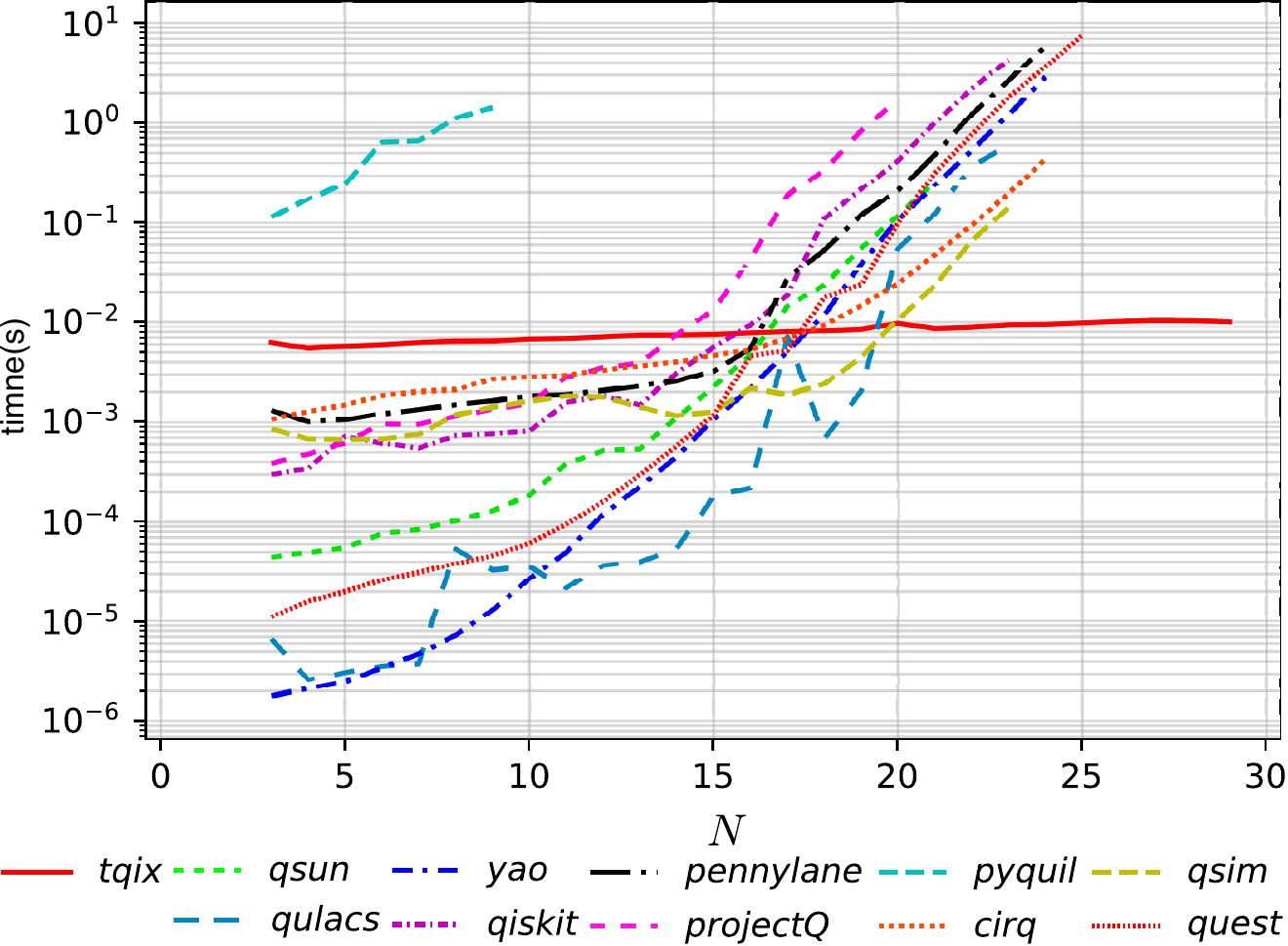}
\caption{
Comparison of the running time between
various libraries. 
The benchmarking circuit is 
the one that generates a GHZ state.
}
\label{fig:4}
\end{figure}

Finally, we compare \texttt{tqix.pis} 
library with other libraries 
including qsim \cite{qsim},
cirq \cite{cirq},
qulacs \cite{Suzuki2021qulacsfast},
yao \cite{Luo2020yaojlextensible},
qsun \cite{Nguyen_2022},
quest \cite{Jones2019},
pennylane \cite{https://doi.org/10.48550/arxiv.1811.04968},
qiskit \cite{qiskit},
projectQ \cite{Steiger2018projectqopensource},
and qypuil \cite{pyquil}.
The results are shown in Fig.~\ref{fig:4}.
We consider a quantum circuit
that generates the maximum entanglement GHZ state.
In \texttt{tqix.pis}, a GHZ state is 
created by applying the GMS$(\pi/2,0)$ gate 
onto the initial ground state. 
In other libraries, the circuit 
encompasses one Hadamard gate
and $N-1$ CNOT gates.
For $N < 30$ as shown in the figure,
\texttt{tqix.pis} 
remains the running time lower than
$10^{-2}$ seconds 
while the time for others 
increases exponentially versus $N$.
See the full code in \ref{appD}.
The \texttt{tqix.pis} represents 
quantum states and operators 
by sparse matrices; 
thus, it consumes a certain running time
and less depend on (small) $N$.
The other libraries represent
quantum states by state vectors,
where the running time 
depends on $2^N$.

Note that in this comparison, we aim to 
emphasize different perspectives 
for different simulation libraries:
while previous libraries use full Hilbert space
to perform general quantum simulation framework,
our library applies the collective Hilbert subspace to
simulate collective phenomena in ensemble systems.
%

\section{Applications}
\label{sec5}
\subsection{Spin squeezing}

\begin{figure*} [t]
\centering
\includegraphics[width=17.6cm]{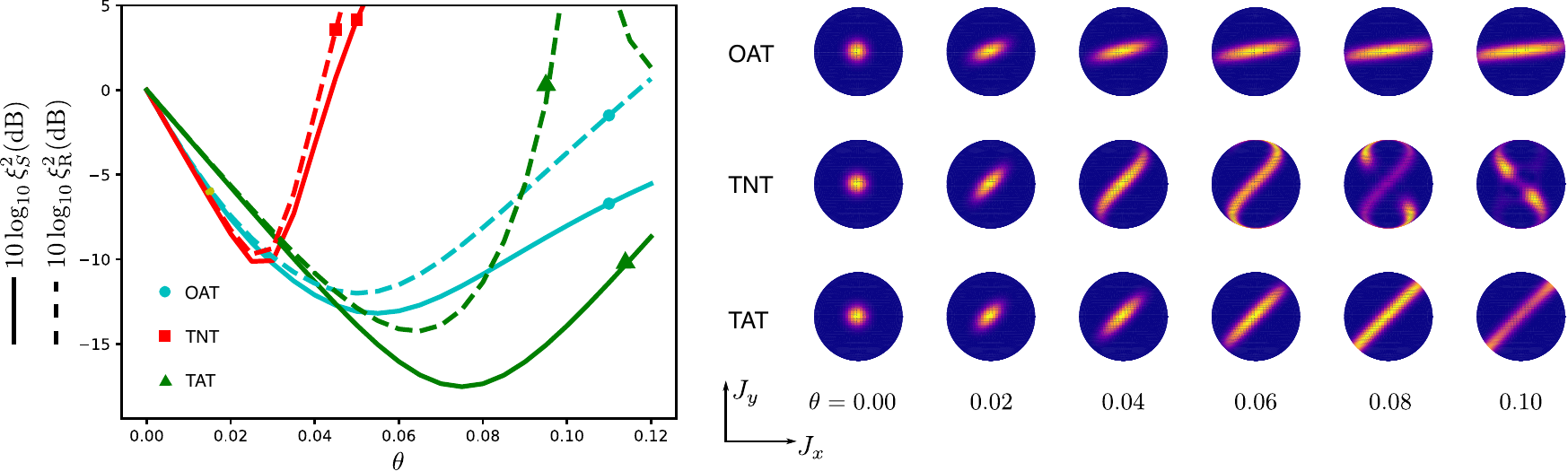}
\caption[20]{(left) Spin squeezing parameters
$\xi^2_S$ and $\xi^2_R$ (in dB) are generated by 
OAT, TNT, and TAT gates.
There exist an optimal $\theta$ 
so that the squeezing reaches the maximum. 
(right) The corresponding Husimi visualization 
for several values of $\theta$.
The figure is plotted at $N = 100$ particles.
}
\label{fig:5}
\end{figure*} 

\textbf{Definition}.
There are several definitions 
for the squeezing parameter in the literature 
\cite{PhysRevA.47.5138,PhysRevA.50.67,Sorensen2001,MA201189}.
%
%
The two well-known
spin squeezing parameters are
given by Kitagawa and Ueda \cite{PhysRevA.47.5138} 
and Wineland \cite{PhysRevA.50.67}.
According to Kitagawa and Ueda,
the spin-squeezing state (SSS) 
redistributes quantum fluctuations 
between two noncommuting observables 
while preserving the minimum uncertainty product
\cite{PhysRevA.78.052101, PhysRevA.81.032104}. 
The corresponding squeezing parameter is given by
\begin{align}\label{eq:xi_S_2}
    \xi^2_{S}=
     \frac{2}{N}
     \Big[\langle \bm J^2_{\bm n_2}+\bm J^2_{\bm n_3}\rangle
     \pm \sqrt{
     \langle \bm J^2_{\bm n_2}-\bm J^2_{\bm n_3}\rangle^2
     +4{\rm cov}^2(\bm J_{\bm n_2},\bm J_{\bm n_3})}
     \Big],
\end{align}
where
\begin{align}
\notag  {\bm n}_2 &= (-\sin\phi,\cos\phi,0);\\
\notag  {\bm n}_3 &= (\cos\theta \cos\phi,
        \cos\theta \sin\phi,-\sin\theta),
\end{align}
and 
\begin{align}
\notag   \theta &= \arccos \left( 
   \frac{\langle J_z\rangle}{|\bm{J}|} 
   \right)
   \text{, } \\
   \phi &=  
        \begin{cases}
            \arccos
            \left( 
            \frac{\langle J_x \rangle}
            {|\bm{J}\sin\theta|} 
            \right)    
   & \text{if } \langle J_y \rangle > 0, \\
   2\pi - \arccos\left( 
   \frac{\langle J_x \rangle}
   {|\bm{J}\sin\theta|} 
   \right) 
   & \text{if } 
   \langle J_y \rangle \leq 0 ,
   \end{cases}
\end{align}
with 
$
  |\bm{J}| = \sqrt{ \langle J_x \rangle^2 +
  \langle J_y \rangle^2 + \langle J_z \rangle^2}.
$
These notations are given 
in the spherical coordinate.
The covariant cov$(\bm J_{\bm n_2},
\bm J_{\bm n_3})$ is given by
\begin{align}\label{eq:cov}
    {\rm cov}(\bm J_{\bm n_2},\bm J_{\bm n_3}) 
    = \frac{1}{2} 
    \langle 
    [\bm J_{\bm n_2},\bm J_{\bm n_3}]_+ 
    \rangle - 
    \langle \bm J_{\bm n_1} \rangle
    \langle \bm J_{\bm n_2} \rangle. 
\end{align}

Similarly, the Wineland squeezing parameter 
is defined by
\cite{PhysRevA.50.67}
\begin{align}\label{eq:xi_R_2}
     \xi^2_{R}=
     \left(\ \frac{N}
     {2|\langle \bm{J} \rangle|}
     \right)^2\xi^2_{S}.
\end{align}
When $\xi^2 < 1$, the system state is squeezed. 

In \texttt{tqix.pis},
to compute the squeezing parameters,
we call \texttt{get\_xi\_2\_S}
and \texttt{get\_xi\_2\_R} functions
for $\xi_S^2$ and $\xi_R^2$, respectively. For example,
we generate a quantum circuit with $N = 50$ 
particles and calculate its squeezing parameters as follows:
\begin{lstlisting}
qc = circuit(N = 50)
xiS2 = get_xi_2_S(qc)
xiR2 = get_xi_2_R(qc)
\end{lstlisting}

\textbf{Spin squeezing by nonlinear gates}.
Spin squeezing state (SSS) can be created
from the CSS by applying a nonlinear gate
such as OAT, TNT, and TAT.
Here, we illustrate the spin squeezing
for these three cases using 
\texttt{tqix.pis} as follows.
First, we apply the rotation gate ${\rm RN}(\pi/2,0)$ 
onto the initial quantum register 
to transform the ground state into the CSS state. 
Then, we apply different nonlinear gates, 
including OAT($\theta)$, TNT($\theta)$ and TAT($\theta, N/2)$. 
We calculate the squeezing parameters of the final states
in these cases and show the result in figure~\ref{fig:5}.
In the left figure, we plot $\xi^2_S$(dB) and $\xi^2_R$(dB).
There exhibit squeezing
for all cases, i.e., $\xi^2$(dB) $< 0$,
and there exists a lower bound for 
each squeezing parameter at a certain $\theta$.
The $\xi^2_R$'s curves are always 
higher than $\xi^2_S$ since $\frac{N}{2} 
\ge |\langle \bm {J} \rangle|$
in Eq.~\eqref{eq:xi_R_2}.
The corresponding Husimi visualizations 
are shown in the right figure 
for several values of $\theta$,
in which displace different types 
of squeezing due to different 
types of nonlinear gate
(OAT, TNT, and TAT).
We plot the results for $N = 100$ particles. 
See the full code in \ref{appE}.

\textbf{Spin squeezing by variational algorithms}.
It can be seen from the previous results, 
these squeezing parameters reflect the presence of an optimal SSS when tuning to the right $\theta$, 
so it is reasonable to use optimization algorithms 
to find optimal circuit's parameters corresponding to the minimum of squeezing function. 
Here, we introduce a quantum-classical 
hybrid variational scheme. 
In the quantum part, we sequentially apply 
a set of OAT($\theta_1$), 
TNT ($\theta_2$), 
and TAT($\theta_3,N/2$), 
into a CSS circuit. 
We then measure the final state 
on the Dicke basis 
and send the results into 
the classical part 
to compute the cost function. 
For generating SSS, 
we apply the cost function to be
the squeezing parameter
\begin{align}\label{eq:cost_function}
    \mathcal{C}(\bm \theta) = 
    \xi^2_{S}(\bm \theta),
\end{align}
where $\bm\theta = (\theta_1, \theta_2, \theta_3)$.
We solve the following optimization problem 
\begin{align}\label{eq:optimize theta}
    \bm \theta^* = \arg\min_{\bm \theta} 
    \mathcal{C}(\bm \theta),
\end{align}
by using gradient-based optimizes including 
gradient descent (GD), 
Adam, and quantum natural gradient descent (QNG)
(see \ref{appF} for the details.)
The gradient of $\mathcal{C}$ with respect 
to $\theta\, \forall \theta \in\bm\theta$ 
is derived by a finite difference (findiff) 
approximation for sparse matrix type
(use in CPUs)
\begin{align}\label{eq:finite difference}
    \frac{d\mathcal{C}}{d\theta_i} \approx 
    \frac{\mathcal{C}(\theta_i+\epsilon) - 
    \mathcal{C}(\theta_i-\epsilon)}{2\epsilon},
\end{align}
which small $\epsilon$,
and by an automatic differential (autodiff)
mechanism for tensor type
(use in GPUs).
Different from findiff, 
the autodiff works on 
the chain rule of the differentials
with two modes, forward accumulation 
and reverse accumulation
\cite{BARTHOLOMEWBIGGS2000171,
doi:10.1137/080743627}.

\begin{figure}[t]
\centering
\includegraphics[width=8.6cm]{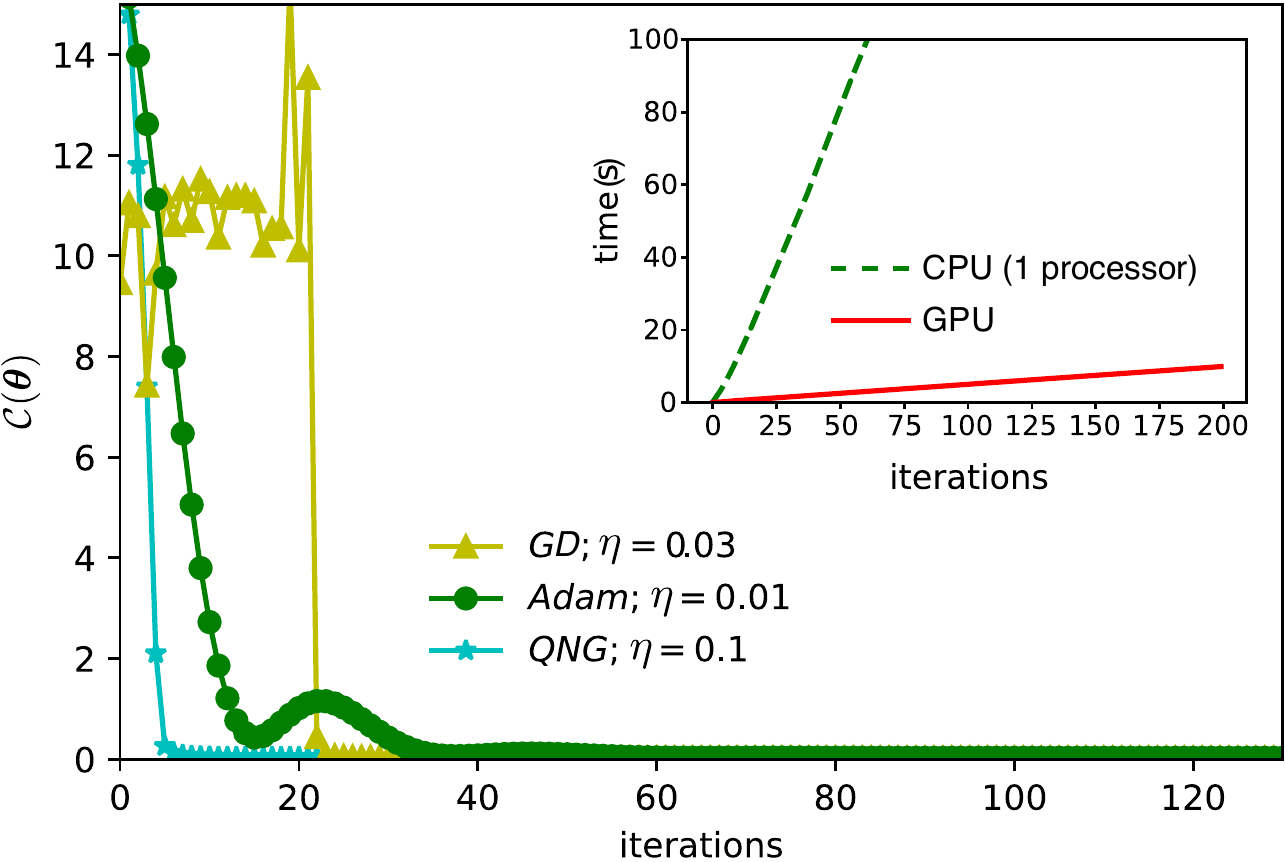}
\caption{
Plot of the cost function versus
the number of iterations 
for different optimizers and learning rates.
Among these optimizers, 
the QNG requires the 
smallest number of iterations
to achieve convergence,
the Adam is not stable near the optimal point,
while the GD is unstable initially but quickly 
drops to the optimal value.
Inset: comparing the execution times between
the CPU and GPU devices.
}
\label{fig:6}
\end{figure}

We choose the initial 
$\bm\theta$ randomly 
and iteratively optimize 
the cost function 
as shown in the results 
in the main figure \ref{fig:6}.
Among the three optimizes, 
the QNG converges fastest 
while the Adam 
results in a less stable 
near the optimal value, 
and the GD fluctuates 
in the first several iterations 
then towards 
the optimal value 
without fluctuating.
This observation reveals the 
natural features of these 
optimizers
\cite{Stokes2020quantumnatural,
https://doi.org/10.48550/arxiv.2107.14063,
https://doi.org/10.48550/arxiv.2204.11635}.
The optimal values 
$\bm\theta^*$ 
and the optimal cost function  
are given in Tab.~\ref{tab:3}. 
The results are plotted at 
$N = 100$, and the 
custom learning rate 
$\eta$ for each optimize
are shown in the main 
figure~\ref{fig:6}. 
See the full code in \ref{appG}.

\begin{table}[!h]
\small
\centering
  \caption{Optimal parameters $\bm \theta$ and its respective cost values for each optimizer.}
  \begin{tabular}{|K{1.3cm}|K{1.3cm}|K{1.3cm}|K{1.3cm}|K{1.3cm}|}
    \hline
        Algorithm & $\theta_1$ & $\theta_2$ &  $\theta_3$ & $\mathcal{C}(\bm \theta)$ \\
    \hline
    GD  & -0.06292 & 0.07942 & -0.02455 & 0.02273 \\
    \hline
    ADAM & -0.03632 & 0.10609 & 0.00115 & 0.02622 \\
    \hline
    QNG & -0.08166 & 0.11887 & 0.01525 & 0.03895 \\
    \hline 
    \end{tabular}
  \label{tab:3}
\end{table}

In addition, we compare the running time 
through the iteration 
for two types of devices: CPU and GPU. 
It can be seen from the inset 
figure~\ref{fig:6} that the GPU 
runs faster than the CPU. 
The reason is that the autodiff 
in GPUs is more optimized 
than the findiff approximation in CPUs. 
The autodiff only calculates 
the cost function once 
and uses the reverse accumulation 
for calculating the gradient. 
Whereas the findiff needs to calculate 
the partial derivative 
for every $\theta_i$, 
where for each $\theta_i$, 
the cost function is computed twice: 
once for $\theta_i+\epsilon$ 
and once for $\theta_i-\epsilon$ 
(see Eq.~\eqref{eq:finite difference}.)
Moreover, the findiff is difficult 
to convert, leading to inefficient code
and intractable to perform higher 
order derivatives resulting in complexity
and high error rate. 
The autodiff, however, 
can overcome these limitations. 

\subsection{Quantum phase transition}
Quantum phase transition (QPT) is a transition between different quantum phases, such as the change in the ground state phases under the variation of magnetic fields
(see \cite{sachdev_2011} for reference.) 
Let us consider the 
Lipkin-Meshkov-Glick (LMG) model 
for an ensemble spins system
interacting through a spin-spin infinite-range exchange coupling $\lambda$ and exposing under an effective transverse field $h$.
The interaction Hamiltonian is given by \cite{LIPKIN1965188}
\begin{align}\label{eq:HLMG}
    H_{\rm LMG} =
    -2hJ_z - \frac{2\lambda}{N}
    (J_x^2 - J_y^2).
\end{align}
Due to the spin-spin interaction,
the system may exhibit 
the quantum phase transition
under the adiabatic dynamics
\cite{PhysRevB.71.224420,PhysRevB.78.104426,PhysRevLett.112.030403}. 

We illustrate the QPT using \texttt{tqix.pis}
library. Let us set the ratio $r = h/\lambda$ for a fixed $\lambda$,
and recast Eq.~\eqref{eq:HLMG} as
\begin{align}\label{eq:HLMGr}
    H_{\rm LMG} = -2\lambda
     \Big[rJ_z + \textstyle\frac{1}{N}
    (J_x^2 - J_y^2)\Big].
\end{align}
Its unitary evolution is given
in terms of quantum gates as
\begin{align}\label{eq:ULMGr}
\notag    U_{\rm LMG} = e^{-itH_{\rm LHG}}
    &\approx e^{-i\lambda' rJ_z}
    e^{-i\frac{\lambda'}{N} (J_x^2-J_y^2)}\\
    & = {\rm RZ}(\Lambda r)\
    {\rm TAT}(\Lambda/N, 
    \text{\squote} xy\text{\squote} ),
\end{align}
where we ignored higher order terms 
in $t$, i.e., the interaction time is short,
and used $\Lambda = -2t\lambda$.

\begin{figure}[t]
\centering
\includegraphics[width=8.8cm]{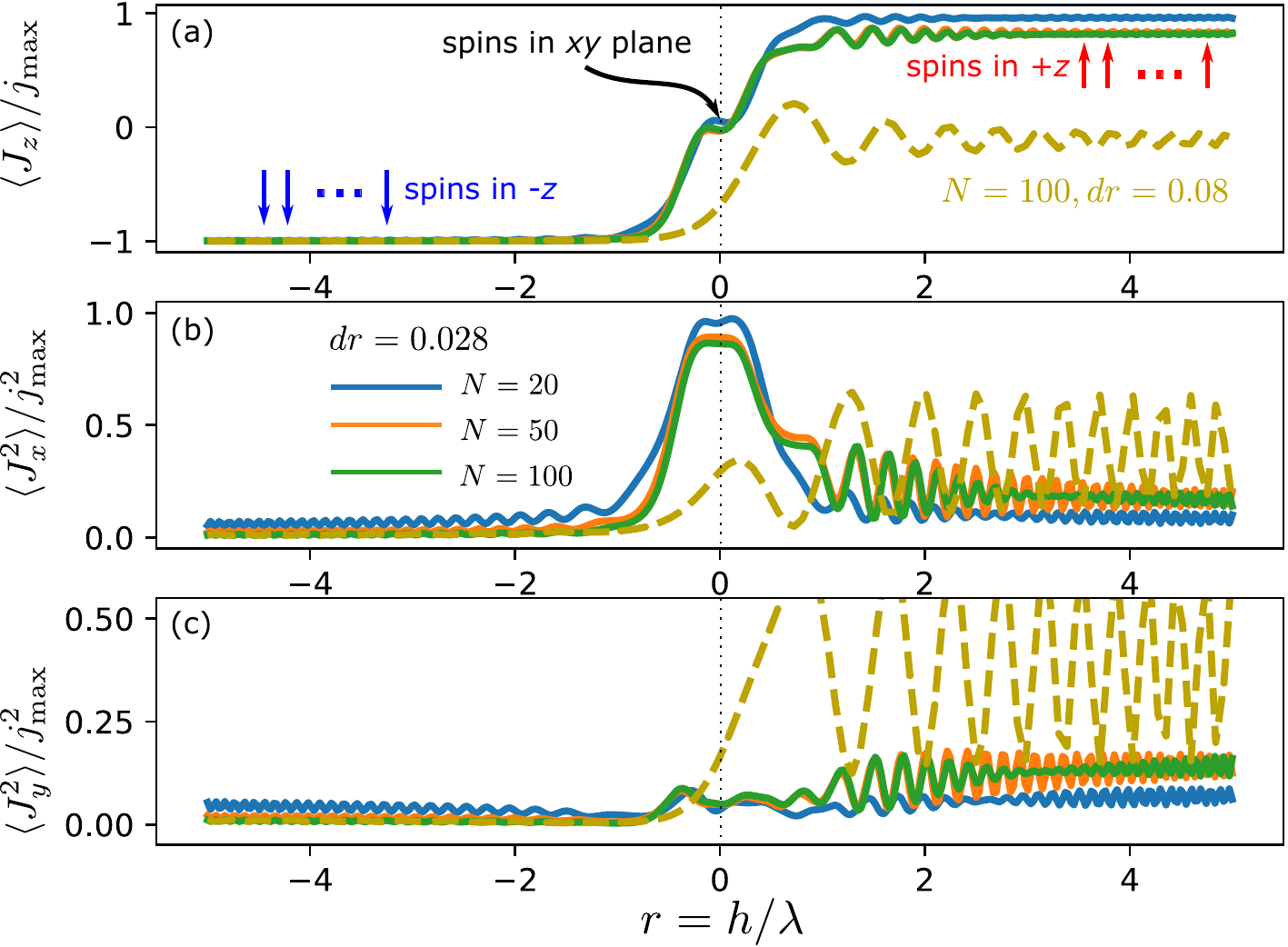}
\caption{
Quantum phase transition in 
the ensemble spins system
via the change in expectation values
when $h$ changes its sign from minus to plus. 
}
\label{fig:7}
\end{figure}

The results are shown in 
Fig.~\ref{fig:7} for the first- 
and second-order of 
the expectation values,
i.e., $\langle J_z\rangle$,
$\langle J_x^2\rangle$, and 
$\langle J_y^2\rangle$.
We fixed $\Lambda = -0.2$.
We observe 
the phase transition as 
the change in these 
expectation values
when the external 
transverse field $h$ 
changes its sign 
(the solid curves,) 
which is similar to
the previous study
\cite{PhysRevLett.100.040403}.
For example, in figure (a), 
starting from a negative 
$h < 0$,
all the spins align 
along the $-z$ direction,
i.e., the ground state dominated 
by the transverse field. 
Adiabatically changing 
$h$ to the critical point, 
i.e., $h = 0$, 
then all the spins order 
in the $xy$ plane. 
Continuing increasing 
$h$ up to
positive, all the spins 
align along the $+z$ 
direction, i.e., the
phase completely changes.

To observe the QPT, 
we must choose the appropriate 
annealing time via $r$-step
($dr$),
according to the adiabatic 
annealing evolution
\cite{doi:10.1126/science.1057726,doi:10.1137/080734479}.
The annealing time $dr$ is defined by
$dr = \frac{r_{\rm max} - r_{\rm min}}{T}$,
$T$ is the total time.
We used $dr = 0.028$ (solid curves) 
in the above example 
and observed the QPT. 
If the annealing time is larger, 
e.g., $dr = 0.08$ (dashed curves), 
then the expectation values 
oscillate after the critical point. 
Detailed code
is given in \ref{appH}.
 
\section{Conclusion}
\label{sec6}
We developed 
\texttt{tqix.pis},
an object-oriented library 
for quantum dynamics simulation
of spin ensemble in Dicke basis.
We applied the collective process 
in the ensemble of psin-1/2 particles and thus 
reduced the dimension of the whole system.
The library allows for simulating
quantum dynamics with collective states
and collective operators. 
Furthermore, it integrates parallelizing 
multi-core processors 
and Graphics Processing Units,
making it run faster. 
We finally showed two representative applications
on quantum squeezing and quantum phase transition. 
Our library is a practical tool 
for simulation collective phenomena 
in ensemble systems with 
low computational cost, 
and we look forward near future 
applications in many-body 
quantum dynamics.

\section*{Declaration of competing interest}
The authors declare that they have no known 
competing financial interests or personal relationships 
that could have appeared to influence 
the work reported in this paper.

\section*{Acknowledgment}
This work is supported by the Ensemble Grants 
for Early Career Researches in Tohoku University.

\section*{Author contributions}
\textbf{Nguyen Tan Viet}: wrote and implemented the code 
and performed the numerical calculations.
\textbf{Nguyen Thi Chuong}: benchmarked 
and compared various libraries. 
\textbf{Vu Thi Ngoc Huyen}: proposed and analyzed the application on quantum phase transition.
\textbf{Le Bin Ho}: proposed the theoretical framework, 
designed the model, 
and scrutinized the results.  
All authors discussed and contributed to the final manuscript.

\appendix

\section{An example code for 
employing a quantum circuit in \texttt{qutip.pis}}
\label{appA}
In \Colorbox{bkgd}{\texttt{tqix.pis}}
to create a quantum circuit,
we call 
\Colorbox{bkgd}{\texttt{circuit}} function,
where \texttt{circuit.state} is a sparse 
matrix representation for the initial quantum state. 
The action of quantum gates on the state
is given by \texttt{circuit.gate\_name},
while the measurement is executed
by \texttt{circuit.measure(num\_shots)}.

\begin{lstlisting}
from tqix import *
from tqix.pis import *
import numpy as np

numq=50

#call the initial circuit
qc = circuit(numq)
psi = qc.state
#print psi
print(psi) #sparse matrix
print(psi.toarray()) #full matrix

#apply the rotation gate RN on the circuit
qc.RN(-np.pi/2,np.pi/4)
psi2 = qc.state

#visualize state
THETA = [0, np.pi]
PHI = [0, 2* np.pi]
husimi_spin_3d(psi.toarray()+psi2.toarray(),THETA ,PHI,cmap = cmindex(1),dirname="./FIG",fname ="husimi_sphere.eps",view=(0,0))

#get probability
prob = qc.measure(num_shots=1000)
#plot figure
from matplotlib import pyplot as plt
x = np.arange(0,numq+1,1)
plt.bar(x,prob)
plt.savefig("./FIG/Pjm.eps")
\end{lstlisting}

\section{Full code for producing Fig.~\ref{fig:3} in the main text}
\label{appB}
\begin{lstlisting}
from tqix.pis import *
from tqix import *
import time 
import matplotlib.pyplot as plt

N_max=200 #max number of particles
L = 3 #number of layers 

#run time benchmark on noiseless system
noiseless_time_qubits = []
for N in range(1,N_max):
    qc = circuit(N)

    start = time.time()

    for _ in range(L):
        qc.RX(np.pi/3)
        qc.RY(np.pi/3)
        qc.RZ(np.pi/3)

    noiseless_time = time.time()-start
    noiseless_time_qubits.append(noiseless_time)

#run time benchmark on noise system with #1 process for simulating noise
noise_time_qubits = []
for N in range(1,N_max):
    qc = circuit(N)

    start = time.time()

    for _ in range(L):
        qc.RX(np.pi/3,noise=0.05)
        qc.RY(np.pi/3,noise=0.05)
        qc.RZ(np.pi/3,noise=0.05)

    noise_time = time.time()-start
    noise_time_qubits.append(noise_time)


#run time benchmark on noise system with #multi-processes for simulating noise
mp_noise_time_qubits = []
for N in range(1,N_max):
    qc = circuit(N,num_process=25)

    start = time.time()

    for _ in range(L):
        qc.RX(np.pi/3,noise=0.05)
        qc.RY(np.pi/3,noise=0.05)
        qc.RZ(np.pi/3,noise=0.05)

    mp_noise_time = time.time()-start
    mp_noise_time_qubits.append(mp_noise_time)

#run time benchmark on noise system with #gpu for simulating noise
gpu_noise_time_qubits = []
for N in range(1,N_max):
    qc = circuit(N,use_gpu=True)

    start = time.time()

    for _ in range(L):
        qc.RX(np.pi/3,noise=0.05)
        qc.RY(np.pi/3,noise=0.05)
        qc.RZ(np.pi/3,noise=0.05)

    gpu_noise_time = time.time()-start
    gpu_noise_time_qubits.append(gpu_noise_time)

#plot data
plt.plot(range(1,len(noiseless_time_qubits)+1), noiseless_time_qubits, label = "noiseless")
plt.plot(range(1,len(noise_time_qubits)+1), noise_time_qubits, label = "noise (#process = 1)")
plt.plot(range(1,len(mp_noise_time_qubits)+1), mp_noise_time_qubits, label = "noise (#process = 25)")
plt.plot(range(1,len(gpu_noise_time_qubits)+1), gpu_noise_time_qubits, label = "noise (GPU: NVIDIA V100)")
plt.legend()
plt.show()
plt.savefig("figtimebm.eps")
\end{lstlisting}

\section{Time complexity in a quantum gate execution}
\label{appC}
We categorize four cases of 
acting quantum gates on
states and compute the 
time complexities $\mathcal{O}$
(big O) as follows.
We consider the 
``worse case" for every evaluation.

+ Case 1: \textbf{Noiseless and symmetry state}.
The complexity for creating an 
operator $J$ is $\mathcal{O}(N^2)$. The
complexity for the state evolution with $e^{-itJ}$ is $\mathcal{O}(N^3)$. The complexity for
computing the density state in Eq.~\eqref{eq:evoled_state} is $\mathcal{O}(N^2)$. In total, the time complexity for this case is $\mathcal{O}(N^3)$.

+ Case 2: \textbf{Noiseless and collective state}.
In the collective form, the cost of creating $J$ is the total cost of creating block matrices (as in Fig.~\ref{fig:1}), each has $\mathcal{O}(i^2)$ with $i \in [j_{min}, \frac{N}{2}$], so creating $J$ costs $\mathcal{O}(N^3)$.
The evolution of the state with $e^{-itJ}$ costs $\mathcal{O}((d^j_N)^3)=\mathcal{O}(N^6)$. Therefore, the time complexity for this case is $\mathcal{O}(N^6)$.

+ Case 3: \textbf{Noise and symmetry state}.
The first stage in this case 
is the same as case 1, 
so the time complexity is 
$\mathcal{O}(N^3)$. When adding noise, 
we have to process $\mathcal{O}(N^2)$ 
elements in the block-$\frac{N}{2}$. 
Each element costs 
$\mathcal{O}((d^j_N)^2)
=\mathcal{O}(N^4)$. In total, 
the time complexity 
of the noise process is 
$\mathcal{O}(N^4N^2) 
= \mathcal{O}(N^6)$.

+ Case 4: \textbf{Noise and collective state}.
The procedure is the same as case 3 
but we have to calculate all elements 
that are non-zeros in all block matrices, 
we have $\mathcal{O}(N^3)$ elements, 
each element costs $\mathcal{O}(N^4)$. 
Therefore, the time complexity 
is $\mathcal{O}(N^7)$.


\section{Full code for producing Fig.~\ref{fig:4} in the main text}
\label{appD}
\begin{lstlisting}
from matplotlib import pyplot as plt
import numpy as np
import pickle

lb_res = pickle.load(open("./lib_benc_data.pickle","rb")) #load time data benchmarked # for libraries  
ax = plt.gca() 
ax.plot(range(1,len(lb_res["tqix"])+1),np.log(lb_res["tqix"]),'b-*',label=r'$tqix$')
ax.plot(range(1,len(lb_res["qsun"])+1),np.log(lb_res["qsun"]),'g--^',label=r'$qsun$')
ax.plot(range(1,len(lb_res["qulacs"])+1),np.log(lb_res["qulacs"]),'r--o',label=r'$qulacs$')
ax.plot(range(1,len(lb_res["yao"])+1),np.log(lb_res["yao"]),'c-.',label=r'$yao$')
ax.plot(range(1,len(lb_res["qiskit"])+1),np.log(lb_res["qiskit"]),'m-+',label=r'$qiskit$')
ax.plot(range(1,len(lb_res["pennylane"])+1),np.log(lb_res["pennylane"]),'y--x',label=r'$pennylane$')
ax.plot(range(1,len(lb_res["projectQ"])+1),np.log(lb_res["projectQ"]),'k-1',label=r'$projectQ$')
ax.plot(range(1,len(lb_res["pyquil"])+1),np.log(lb_res["pyquil"]),'c--v',label=r'$pyquil$')
ax.plot(range(1,len(lb_res["cirq"])+1),np.log(lb_res["cirq"]),'g->',label=r'$cirq$')
ax.plot(range(1,len(lb_res["qsim"])+1),np.log(lb_res["qsim"]),'y-3',label=r'$qsim$')
ax.plot(range(1,len(lb_res["quest"])+1),np.log(lb_res["quest"]),'m-s',label=r'$quest$')

ax.set_xlabel("number of particles")
ax.set_ylabel("time execution (log scale)")
lgd = ax.legend(loc='center left', bbox_to_anchor=(1, 0.5))
plt.savefig("./lib_benchmark.eps", bbox_extra_artists=(lgd,), bbox_inches='tight')

\end{lstlisting}

\section{Full code for producing Fig.~\ref{fig:5} in the main text}
\label{appE}
\begin{lstlisting}
from tqix.pis import *
from tqix import *
import numpy as np
from matplotlib import pyplot as plt
import os
N=100
THETA = [0, np.pi]
PHI = [0, 2* np.pi]
angles = np.linspace(0,0.5,100).tolist()

if os.path.isdir("./OAT"):
      pass 
else:
      os.mkdir("./OAT")

if os.path.isdir("./TNT"):
      pass 
else:
      os.mkdir("./TNT")

if os.path.isdir("./GMS"):
      pass 
else:
      os.mkdir("./GMS")

if os.path.isdir("./TAT"):
      pass 
else:
      os.mkdir("./TAT")

# OAT
OAT_xi_2_S = []
OAT_xi_2_R = []
for theta in angles:
      qc = circuit(N)
      qc.RN(np.pi/2,0)
      qc.OAT(theta,"Z")
      OAT_xi_2_S.append(10*np.log10(np.real(get_xi_2_S(qc))))
      OAT_xi_2_R.append(10*np.log10(np.real(get_xi_2_R(qc))))

# plot sphere 
for theta in ([0.0, 0.02, 0.04, 0.06, 0.08,0.1]):
     qc = circuit(N)
     qc.RN(np.pi/2,0)
     qc.OAT(theta,"Z")
     husimi_spin_3d(qc.state.toarray(), THETA ,PHI ,cmap = cmindex(1),dirname="./OAT",fname =str(theta)+"husimi_sphere.eps",view=(180,0))

#TNT
TNT_xi_2_S = []
TNT_xi_2_R = []

for theta in angles:
      qc = circuit(N)
      qc.RN(np.pi/2,0)
      omega = N*theta
      qc.TNT(theta,omega=omega,gate_type="ZX")
      TNT_xi_2_S.append(10*np.log10(np.real(get_xi_2_S(qc))))
      TNT_xi_2_R.append(10*np.log10(np.real(get_xi_2_R(qc))))

# plot sphere 
for theta in ([0.0, 0.02, 0.04, 0.06, 0.08,0.1]):
     qc = circuit(N)
     qc.RN(np.pi/2,0)
     qc.TNT(theta,"ZX")
     husimi_spin_3d(qc.state.toarray(), THETA ,PHI ,cmap = cmindex(1),dirname="./TNT",fname =str(theta)+"husimi_sphere.eps",alpha=1,view=(180,0))
     
#GMS
GMS_xi_2_S = []
GMS_xi_2_R = []
phi = np.pi/4 
for theta in angles:
      qc = circuit(N)
      qc.GMS(theta,phi)
      GMS_xi_2_S.append(10*np.log10(np.real(get_xi_2_S(qc))))
      GMS_xi_2_R.append(10*np.log10(np.real(get_xi_2_R(qc))))

# plot sphere 
for theta in ([0.0, 0.02, 0.04, 0.06, 0.08,0.1]):
     qc = circuit(N)
     qc.GMS(theta,phi)
     husimi_spin_3d(qc.state.toarray(), THETA ,PHI ,cmap = cmindex(1),dirname="./GMS",fname =str(theta)+"husimi_sphere.eps",view=(-90,0))

# TAT
TAT_xi_2_S = []
TAT_xi_2_R = []
for theta in angles:
       qc = circuit(N)
       qc.RN(np.pi/2,0)
       qc.TAT(theta,"ZY")
       TAT_xi_2_S.append(10*np.log10(np.real(get_xi_2_S(qc))))
       TAT_xi_2_R.append(10*np.log10(np.real(get_xi_2_R(qc))))

# plot sphere 
for theta in ([0.0, 0.02, 0.04, 0.06, 0.08,0.1]):
     qc = circuit(N)
     qc.RN(np.pi/2,0)
     qc.TAT(theta,"ZY")
     husimi_spin_3d(qc.state.toarray(), THETA ,PHI ,cmap = cmindex(1),dirname="./TAT",fname =str(theta)+"husimi_sphere.eps",view=(180,0))

ax = plt.gca() 
ax.plot(angles, OAT_xi_2_S,'c-o',label=r'$10log_{10}(\xi^{2}_{S})-OAT$')
ax.plot(angles, TNT_xi_2_S,'r-s',label=r'$10log_{10}(\xi^{2}_{S})-TNT$')
ax.plot(angles, TAT_xi_2_S,'g-*',label=r'$10log_{10}(\xi^{2}_{S})-TAT$')
ax.plot(angles, GMS_xi_2_S,'y-o',label=r'$10log_{10}(\xi^{2}_{S})-GMS$')
ax.plot(angles, OAT_xi_2_R,'c--o',label=r'$10log_{10}(\xi^{2}_{R})-OAT$')
ax.plot(angles, TNT_xi_2_R,'r--s',label=r'$10log_{10}(\xi^{2}_{R})-TNT$')
ax.plot(angles, TAT_xi_2_R,'g--*',label=r'$10log_{10}(\xi^{2}_{R})-TAT$')
ax.plot(angles, GMS_xi_2_R,'y--o',label=r'$10log_{10}(\xi^{2}_{R})-GMS$')

ax.set_xlabel("theta")
ax.set_ylabel("Db")
lgd = ax.legend(loc='center left', bbox_to_anchor=(1, 0.5))
dirname= ""
fname ="xi_2_S_R_graph.eps"
plt.savefig(os.path.join(dirname,fname))
plt.close()

\end{lstlisting}

\section{Optimizers used in 
variation quantum squeezing circuit}
\label{appF} 
In a variational circuit, 
the quantum part
is parameterized by
$\bm \theta$ that will be 
literately updated by, such as
gradient-based optimizers.
Here, we consider the gradient descent (GD),
Adam \cite{https://doi.org/10.48550/arxiv.1412.6980}, 
and
Quantum natural gradient (QNG) \cite{Stokes2020quantumnatural}
optimizers.
The GD computes new parameters via
\begin{align}\label{eq:sgd}
	\bm{\theta}^{t+1}
	=\bm{\theta}^{t}-\eta\nabla_{\bm\theta}
	\mathcal{C}(\bm\theta),
\end{align}
where 
$\nabla_{\bm\theta}
\mathcal{C}(\bm\theta)$
is the gradient of $\mathcal{C}(\bm\theta)$,
and $\eta$ is the learning rate.
It is simple, but the coverage is low, and 
one must choose a proper learning rate 
to have the best result.
Whereas, the Adam computes new parameters by
\begin{align}\label{eq:adam}
&\bm{\theta}^{t+1}=\bm{\theta}^{t}
-\eta\frac{\hat{m}_{t}}{\sqrt{\hat{v}_{t}} + \epsilon},
\end{align}
where $m_{t}=\beta_{1} m_{t-1}
+\left(1-\beta_{1}\right) 
\nabla_{\bm\theta}\mathcal{C}(\bm\theta), 
v_{t}=\beta_{2} v_{t-1}+(1-\beta_{2}) 
\nabla_{\bm\theta}^2\mathcal{C}(\bm\theta),
\hat{m}_{t}=m_{t} /\left(1-\beta_{1}^{t}\right),
\hat{v}_{t}=v_{t} /\left(1-\beta_{2}^{t}\right),
$
with 
$\eta = 0.01, \beta_1 = 0.8, 
\beta_2 = 0.999$ 
and $\epsilon = 10^{-10}$. 
The Adam optimizer 
automatically adapts 
the learning rate and fast coverage, 
but it is noisy near the optimal point.
Finally, the QNG is better than other optimizers 
but requires more computational cost 
regards to quantum circuits.
It is given by
\begin{align}\label{eq:QuanNat}
    \bm{\theta}^{t+1}=\bm{\theta}^{t}-\eta
    g^+\nabla_{\bm\theta}\mathcal{C}(\bm\theta),
\end{align}
where 
$g^+$ is the pseudo-inverse of a 
Fubini-Study metric tensor $g$ 
\cite{Stokes2020quantumnatural}.
Detailed for deriving the 
Fubini-Study metric tensor
can be seen from Ref.
\cite{https://doi.org/10.48550/arxiv.2204.11635}.


\section{Full code for producing Fig.~\ref{fig:6} in the main text}
\label{appG}
\begin{lstlisting}
from tqix.pis import *
from tqix import *
import numpy as np
from matplotlib import pyplot as plt
import torch 
import numpy as np 

def cost_function(theta,use_gpu=False):
    qc = circuit(N,use_gpu=use_gpu)
    qc.RN(np.pi/2,0)
    qc.OAT(theta[0],"Z") 
    qc.TNT(theta[1],omega=theta[1],gate_type="ZX")
    qc.TAT(theta[2],"ZY")
    if use_gpu:
        loss = torch.real(get_xi_2_S(qc))
    else:
        loss = np.real(get_xi_2_S(qc))
    return  loss

N=100 #number of particles
route = (("RN2",),("OAT","Z"),("TNT","ZX"),("TAT","ZY")) #define layers for QNG 
loss_dict = {}

#function to optimize circuit of sparse array
def sparse(optimizer,loss_dict,mode):
    objective_function = lambda params: cost_function(params) 
    init_params = [0.00195902, 0.14166777, 0.01656466] #random init parameters
    _,_,_,loss_hist,time_iters = fit(objective_function,optimizer,init_params,return_loss_hist=True,return_time_iters = True)
    loss_dict[mode] = loss_hist
    return loss_dict,time_iters

#function to optimize circuit of tensor
def tensor(optimizer,loss_dict,mode):
    objective_function = lambda params: cost_function(params,use_gpu=True) 
    init_params = [0.00195902, 0.14166777, 0.01656466] #random init parameters
    init_params = torch.tensor(init_params).to('cuda').requires_grad_()
    _, _,_,loss_hist,time_iters = fit(objective_function,optimizer,init_params,return_loss_hist=True,return_time_iters = True)
    loss_dict[mode] = loss_hist
    return loss_dict,time_iters

#QNG
optimizer = GD(lr=0.03,eps=1e-10,maxiter=200,use_qng=True,route=route,tol=1e-19,N=N)
loss_dict,_ = tensor(optimizer,loss_dict,"tensor_qng")

#GD
optimizer = GD(lr=0.0001,eps=1e-10,maxiter=200,tol=1e-19,N=N)
loss_dict,_ = tensor(optimizer,loss_dict,"tensor_gd")

#ADAM 
optimizer = ADAM(lr=0.01,eps=1e-10,amsgrad=False,maxiter=200)
loss_dict,sparse_times = sparse(optimizer,loss_dict,"sparse_adam")

optimizer = ADAM(lr=0.001,eps=1e-10,amsgrad=False,maxiter=200)
loss_dict,tensor_times = tensor(optimizer,loss_dict,"tensor_adam")

#plot loss values with respect to iterations
ax = plt.gca() 
ax.plot(range(len(loss_dict["tensor_qng"] )),loss_dict["tensor_qng"] ,'c-*',label=r'$QNG;\eta=0.1$')
ax.plot(range(len(loss_dict["tensor_gd"])),loss_dict["tensor_gd"],'y-^',label=r'$GD;\eta=0.03$')
ax.plot(range(len(loss_dict["tensor_adam"])),loss_dict["tensor_adam"],'g-o',label=r'$Adam;\eta=0.01$')


ax.set_xlabel("iterations")
ax.set_ylabel(r"$C(\theta)$")
ax.set_xlim(0,130)
ax.set_ylim(0,15)
lgd = ax.legend(loc='center left', bbox_to_anchor=(1, 0.25))
plt.savefig("./loss_bm.eps", bbox_extra_artists=(lgd,), bbox_inches='tight')

#for compare running time between using tensor #and sparse array structure, we plot ADAM #as an example

cumsum_vqa_time_res_sparse = 
np.cumsum(loss_dict['sparse_adam'])
cumsum_vqa_time_res_tensor = np.cumsum(loss_dict['tensor_adam'])
ax = plt.gca() 
ax.plot(range(len(cumsum_vqa_time_res_sparse)),cumsum_vqa_time_res_sparse,'g--',label=r'$CPU$')
ax.plot(range(len(cumsum_vqa_time_res_tensor)),cumsum_vqa_time_res_tensor,'r-',label=r'$GPU$')

ax.set_xlabel("iterations")
ax.set_ylabel("time(s)")
ax.set_ylim(0,100)
lgd = ax.legend(loc='upper right')
plt.savefig("./timetensorsparse.eps", bbox_extra_artists=(lgd,), bbox_inches='tight')

\end{lstlisting}

\section{Full code for producing Fig.~\ref{fig:7} in the main text}
\label{appH}
\begin{lstlisting}
from tqix import *
from tqix.pis import *
import numpy as np
from matplotlib import pyplot as plt

N = [20, 50, 100]
figure, axis = plt.subplots(3, 1)
for i in N:
    qc = circuit(i, use_gpu = False)

    aveJz = [] #average J_z
    aveJx2 = [] #average J_x@J_x
    aveJy2 = [] #average J_y@J_y
    
    r_min = -5
    r_max = 5
    r_iter = 357 #dr = (r_min-r_max)/r_iter ~ 0.028
    x = np.linspace(r_min,r_max,r_iter)
    lambda_p = -0.2
    
    for j in x:
        qc.RZ(lambda_p*j)
        qc.TAT(lambda_p/i,'xy')

        tr0 = qc.expval('Jz')/i*2
        tr1 = qc.expval('Jx2')/i**2*4
        tr2 = qc.expval('Jy2')/i**2*4
        
        aveJz.append(np.real(tr0))
        aveJx2.append(np.real(tr1))
        aveJy2.append(np.real(tr2))

    axis[0].plot(x,aveJz)
    axis[1].plot(x,aveJx2)
    axis[2].plot(x,aveJy2)

# change annealing time    
i = 100 #particles
qc1 = circuit(i, use_gpu = False)

aveJz = []
aveJx2 = [] 
aveJy2 = [] 

r_iter = 125 #dr = 0.08
x = np.linspace(r_min,r_max,r_iter)
lambda_p = -0.2

for j in x:
	qc1.RZ(lambda_p*j)
	qc1.TAT(lambda_p/i,'xy')

    tr0 = qc1.expval('Jz')/i*2
    tr1 = qc1.expval('Jx2')/i**2*4
    tr2 = qc1.expval('Jy2')/i**2*4

	aveJz.append(np.real(tr0))
	aveJx2.append(np.real(tr1))
	aveJy2.append(np.real(tr2))

axis[0].plot(x,aveJz, '--')
axis[1].plot(x,aveJx2, '--')
axis[2].plot(x,aveJy2, '--')    
    
plt.savefig("figQPT.eps")
\end{lstlisting}

As mentioned in the main text,
we can observe the QPT by 
choosing the approximate $r$\_step,
from which resolve to the adiabatic annealing evolution. 

\section{Guide for running multi-core and GPU-Acceleration on chip Apple M1}
\label{appI}
It requires MacOS 12.3+ and an ARM Python installation. We can check them by:

\begin{lstlisting}
import platform
platform.platform()

>> macOS-12.4-arm64-arm-64bit (OK)
>> macOS-11.8-x86_64-i386-64bit (NOT)
\end{lstlisting}

Switch to the terminal and create a new ARM environment using Anaconda:

\begin{lstlisting}
CONDA_SUBDIR=osx-arm64 conda create -n ml python=3.9 -c conda-forge
\end{lstlisting}

Modify the CONDA\_SUBDIR variable to permanently use osx-arm64 for future use:
\begin{lstlisting}
conda env config vars set CONDA_SUBDIR=osx-arm64
\end{lstlisting}

From here, we can switch 
between the two 
environments (base) and (ml):

\begin{lstlisting}
conda activate
conda activate ml
\end{lstlisting}

In the (ml) environment,
let's install the required libraries
(including \texttt{tqix} and pyTorch)
and run the codes above normally. 
For more detail, see:
https://towardsdatascience.com/gpu-acceleration-comes-to-pytorch-on-m1-macs-195c399efcc1

\begin{figure}[h]
\centering
\includegraphics[width=8.6cm]{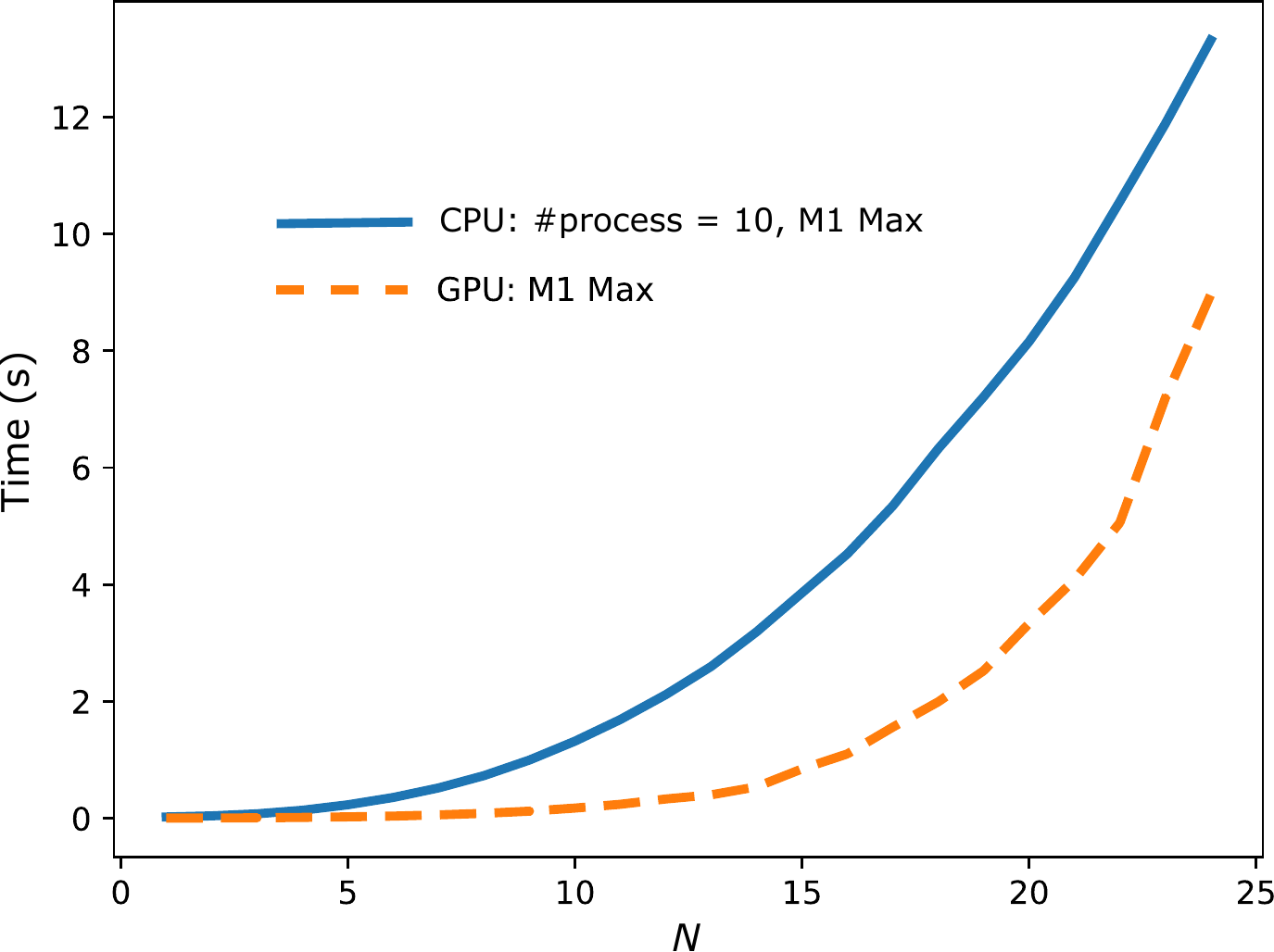}
\caption{
Benchmarking on chip Apple M1 Max
with a comparison between the 10-processor CPU
and the integrated GPU.
}
\label{fig:8}
\end{figure}

We can confirm if the MPS is working
in Python by:
\begin{lstlisting}
import torch
torch.has_mps

>> True
\end{lstlisting}

In Fig~\ref{fig:8},
we compare the running time on chip Apple M1 Max 
with \#process = 10 and GPU.
The system configuration is the same as in figure~\ref{fig:3}.

\bibliographystyle{elsarticle-num}
\bibliography{refs}

\end{document}